\documentclass[AMA,STIX1COL]{WileyNJD-v2}

\newcommand{\norm}[1]{\left\lVert#1\right\rVert}

\articletype{Research Article}%

\usepackage{xr}
\makeatletter
\newcommand*{\addFileDependency}[1]{% argument=file name and extension
  \typeout{(#1)}
  \@addtofilelist{#1}
  \IfFileExists{#1}{}{\typeout{No file #1.}}
}
\makeatother

\newcommand*{\myexternaldocument}[1]{%
    \externaldocument{#1}%
    \addFileDependency{#1.tex}%
    \addFileDependency{#1.aux}%
}
\myexternaldocument{supplement}

\makeatletter
\renewcommand*{\thetable}{\arabic{table}}
\renewcommand*{\thefigure}{\arabic{figure}} 
\makeatother

\begin{document}

\title{Contrasting pre-vaccine COVID-19 waves in Italy through Functional Data Analysis}

\author[1]{Tobia Boschi†}
\author[2]{Jacopo Di Iorio†}
\author[3]{Lorenzo Testa†}
\author[4,5]{Marzia A. Cremona*}
\author[2,3]{Francesca Chiaromonte*}

\authormark{\textsc{}}

\address[1]{IBM Research Europe, Dublin, Ireland}
\address[2]{Department of Statistics, Penn State University, University Park, Pennsylvania (US)}
\address[3]{L'EMbeDS, Sant'Anna School of Advanced Studies, Pisa, Italy}
\address[4]{Department of Operations and Decision Systems, Université Laval, Québec, Canada}
\address[5]{CHU de Québec – Université Laval Research Center, Québec, Canada}

\corres{*Marzia A.~Cremona (marzia.cremona@fsa.ulaval.ca); Francesca Chiaromonte (fxc11@psu.edu)}

%\presentaddress{\lorenzo{Che cosa è supposto essere questo campo?}}

\abstract[Summary]{We use data from 107 Italian provinces to characterize and compare mortality patterns in the first two COVID-19 epidemic waves, which occurred prior to the introduction of vaccines. We also associate these patterns with mobility, timing of government restrictions, and socio‐demographic, infrastructural, and environmental covariates. Notwithstanding limitations in accuracy and reliability of publicly available data, we are able to exploit information in curves and shapes through Functional Data Analysis techniques. Specifically, we document differences in magnitude and variability between the two waves; while both were characterized by a co-occurrence of ``exponential'' and ``mild'' mortality patterns,
%FC: WE SHOULD NOT SAY 2 IN THE ABSTRACT, BECAUSE THROUGH THE MAIN WE DOCUMENT 3 (CLUSTERING), THEN USE 2 (IN THE REGRESSIONS)... WE SHOULD STAY VAGUER UPFRONT!
%two mortality patterns
%(an ``exponential" and a ``milder" one), 
the second spread much more broadly and asynchronously through the country. Moreover, we find evidence of a significant positive association between local mobility and mortality in both epidemic waves, and corroborate the effectiveness of timely
%mobility 
%FC: TAKING OUT THE WORD MOBILITY THIS STAYS MORE GENERAL
restrictions in curbing mortality. The techniques we describe could capture additional signals of interest if applied, for instance, to data on cases and positivity rates. However, we show that the quality of such data, at least in the case of Italian provinces, was too poor to support meaningful analyses.}


% \jnlcitation{\cname{%
% \author{Boschi T.},
% \author{J. Di Iorio},
% \author{L. Testa},
% \author{M.A. Cremona}, and
% \author{F. Chiaromonte}} (\cyear{2023}),
% \ctitle{Contrasting pre-vaccine COVID-19 waves in Italy through Functional Data Analysis}, \cjournal{XXX}, \cvol{2017;00:1--6}.}

\maketitle

\let\thefootnote\relax\footnotetext{† These authors contributed equally}

\section{Introduction}
On January 30, 2020, Italy recorded the first two cases of SARS-Cov-2; a couple of Chinese tourists visiting Rome. 
The first non-travel related case was recorded in the municipality of Codogno (Lombardia) on February 19, 2020. 
With outbreaks quickly emerging in Lombardia and Veneto, local lockdowns were imposed in ten municipalities in the province of Lodi, and in one municipality in the province of Padova -- affecting a total of over $50,000$ people. 
As cases and deaths mounted, local and central authorities took progressively more stringent measures to limit mobility and social gatherings, e.g., suspending university activities, educational trips, sporting events, and exhibitions.
These restrictions culminated with a general nationwide lockdown (March 9, 2020) and the suspension of all non-essential production activities (March 23, 2020), and were not lifted until May -- when free movement was restored within the whole national territory.

After a fairly quiet summer, many countries in Europe, including Italy, witnessed a new rise in COVID-19 cases and deaths, leading to new containment measures. 
Face masks became mandatory, and gathering places such as gyms, theaters, and cinemas closed again. 
On November 4, 2020, in order to control the spread of the epidemic, the Italian government created a classification system whereby regions could be labeled as yellow, orange, or red based on a number of indicators of epidemic spread and stress on the healthcare infrastructure. Each color corresponded to different curfew rules and limitations on mobility. As a precautionary measure, during the holidays red-zone rules were imposed in all regions -- including those where the epidemic was not raging at the time (these were the strictest rules, equivalent to the lockdown of the first wave). In January 2021, as the second wave started to subside, restrictions were loosened again to the pre-holiday color system. 

Prior analyses by our group, which employed Functional Data Analysis (FDA) tools\cite{kokoszka2017, ramsay2005} on data from the first COVID-19 wave at the resolution of regions, characterized different epidemic patterns unfolding in different areas of the country, and suggested a relevant role for mobility and some socio-demographic, infrastructural and environmental factors as statistical predictors of mortality \citep{boschi2021functional}.
Our results were consistent with those of many other studies \cite{collazos2023modeling, engle2020staying,  nepomuceno2020besides, nouvellet2021reduction}. 
In this article, we analyze data from both the first and the second wave of the epidemic -- focusing on the two 150-day intervals from February 25, 2020, to July 23, 2020, and from October 1, 2020, to February 27, 2021. We also increase spatial resolution considering 107 Italian provinces, as opposed to the 20 Italian regions we considered %\lorenzo{before}
in our prior study \cite{boschi2021functional}. 
This ought to prevent
%the 
dilution of signals due to aggregation, and allow for a more detailed characterization of the epidemic and its associations to potential predictors -- in particular, it allows us to employ more complex statistical models without overfitting \citep{hastie2009elements}. Crucially, while very different in terms of restrictions regimes imposed on the population and of its mobility behaviors, the first and second waves of COVID-19 in Italy were both pre-vaccines (the vaccination campaign in the European Union, including Italy, started -- for healthcare professionals and selected fragile categories -- on December 27, 2020). Thus, contrasting them offers an opportunity to refine our understanding of the roles of restrictions and mobility -- without having to account for the effects of the subsequent vaccination campaigns. 

At the outset of our analyses, and yet among our key findings, is evidence of poor data quality. 
Specifically, we observe vast under-counting of deaths in the official reports, especially during the first epidemic wave \citep{ciminelli2020covid, henry2022variation}, and alarming discrepancies in case counts. 
Ignoring information on cases, properly processing information on deaths, and using FDA tools, we find marked differences between the two waves.   
The first was more dramatic, more concentrated spatially -- hitting particularly hard in a handful of provinces in Lombardia, and with relatively homogeneous timing across such provinces. 
In contrast, the second wave was less dramatic but more widespread -- affecting more locations with less homogeneity in timing.
Notably, provinces hardest hit during the first wave were among the least affected during the second; the contrast is particularly striking in the case of Bergamo, which experienced the heaviest mortality during the first wave, making headlines around the world \citep{buoro2020papa, senni2020covid}, and the lightest during the second. 
This may be due to the fact that inhabitants of provinces hardest hit during the first wave adhered more diligently to restrictions and safety mandates during the second, to a reduced number of people at risk (as an effect of the large number of deaths among the most vulnerable individuals during the first wave), or possibly to some kind of herd immunity (as an effect of the large number of cases during the first wave) \cite{golinelli2021small, perico2021bergamo, vinceti2021association}.

To evaluate how the timing of restrictions may affect epidemic unfolding, and more specifically whether introducing restrictions early enough may help curb mortality, we create an \textit{ad-hoc} variable measuring, for each province and separately for the two waves, the area under the mortality curve 
%of a province 
up to the point in time when restrictions were imposed. 
%\lorenzo{With this variable, we aim at capturing the wave-specific cumulative mortality at the time of restriction, thus providing a comprehensive measure of the overall mortality associated with the first stage of each wave, and taking into account the duration and intensity of the epidemic prior to the implementation of restrictive measures.}
Such cumulative mortality should capture how long and how much the epidemic had a chance to ``build up'' in the province prior to the implementation of restrictive measures.
In our analyses, this variable is far stronger than socio-demographic, infrastructural, and environmental factors as a predictor of the mortality unfolding henceforth -- a result consistent with findings in other recent studies \citep{basellini2021explaining, cintia2020relationship}. Assessing and comparing the role of mobility during the two waves is less straightforward. 
Restriction policies were very different in the two waves (uniform across the country in the first, localized and linked to the evolution of epidemic and stress parameters in the second), and so were the behaviors of the population in response to them. 
Likely as a consequence of these differences, the variability observed within the mobility curves of any given province, as well as the variability observed across such curves, changed markedly between the two waves.  
Notwithstanding these changes, we detect a strong and significant positive association between mortality and mobility in both waves -- highlighting the importance of restrictions imposed specifically on the latter as tools to control the epidemic.

The remainder of the article is organized as follows. 
Section \ref{sec:data} describes data, sources, and preprocessing procedures. 
Section \ref{sec:results} presents our main results. First, we show mismatches in case and death counts across different sources and geographical resolutions. Next, we characterize mortality patterns during the two waves and evaluate the effects of various potential predictors -- including socio-demographic, infrastructural, and environmental factors.
Last, we study mobility curves, their dynamics, and their association with mortality curves during the two waves. 
Section \ref{sec:discussion} provides concluding remarks. 

\section{Data and preprocessing}
\label{sec:data}

\subsection{Data on deaths and cases}
\label{subsec:data_deaths}
Unlike our previous work\cite{boschi2021functional}, where we analyzed Italian COVID-19 mortality at the spatial resolution of regions, here we target the higher resolution of provinces. However, to date, data on COVID-19 deaths by province are still not provided by the Italian authorities. 
%\lorenzo{This is still true. See: https://github.com/pcm-dpc/COVID-19/tree/master --- data on provinces only show positives}
We therefore need to resort to \textit{differential mortality}, which can be computed using the daily all-cause death counts provided by the Italian National Statistical Institute (ISTAT) for 7270 Italian municipalities (comprising about 93.5\% of the Italian population). 
In particular, we compute 
%the 
differential mortality as the difference between daily deaths and the average daily deaths in the five years prior to the pandemic (2015--2019), divided by the total population as of January 1, 2019. 
We compute this differential mortality for each province and for each day in the first and the second waves of the epidemic -- in particular in the two 150-day intervals from February 25, 2020, to July 23, 2020, and from October 1, 2020, to February 27, 2021, respectively. 

Province-level differential mortality data are smoothed into curves using cubic \textit{smoothing B-splines} with knots approximately at each week (21 knots equally spaced over 150 days -- the time domain of each wave) and a roughness penalty on the curve second derivative \citep{ramsay2005}. 
Employing one knot per week allows us to focus on general trends and smooth out daily fluctuations. 
% as well as differences between weekdays and weekends. 
% \marzia{[This is certainty true for official DPC data, but I doubt it is true for ISTAT data (I do not remember seeing this in the region data for the first wave)]} \lorenzo{You are absolutely right! Thank you for pointing this out!}
Separately for each wave, the smoothing parameter is selected as to minimize the average generalized cross-validation (GCV) error \citep{craven1978smoothing} across all 107 province-level curves (the same criterion is employed to smooth the other functional data sets in this study; see below). We then align the differential mortality curves using landmark registration \citep{ramsay2005}. Specifically, for each wave, we shift curves horizontally in order to
%make 
align their peaks.
%aligned. 
For curves with multiple peaks, we consider the highest peak occurring between day $10$ and day $100$.
%one in the portion of domain from day 10 to day 100. 
The new common peak time is set to be equal to the time of the earliest peak, specifically day $20$ (peak of the province of Lodi) for the first wave, and day $33$ (peak of the province of Cagliari) for the second wave. 
Finally, we add (remove) days at the end (beginning) of the domain for the shifted curves, to guarantee the same 150-day length for all curves
%that all considered curves have the same length of 150 days
(curves are smoothed again after this ``domain integration'' -- choosing the smoothing parameter by minimizing the GCV criterion). As a result, the common time in each wave is the one of the province with the earliest peak. 
All computations are performed using the \texttt{R} package \texttt{fda} \citep{ramsayfdapackage}.

We repeat mortality calculations
%the calculation 
at the level of regions, in order to compare results with the data provided by the Italian Civil Protection agency (Dipartimento della Protezione Civile; DPC), which releases regional daily counts of recorded COVID-19 deaths since February 2020.  
We divide both ISTAT differential death counts and DPC death counts by the total population as of January 1, 2019, to obtain ISTAT differential mortality and DPC mortality, respectively. 
We do not perform any further preprocessing, since we employ these data only for validation purposes.
DPC also provides official daily counts of COVID-19 cases since February 2020 -- both at the resolutions of provinces \citep{dpc_province} and regions \citep{dpc_region}. However, DPC cases data present several inconsistencies (see Section \ref{subsec:inconsistencies}), hence we do not employ them for further analyses.

\subsection{Socio-demographic, infrastructural, and environmental data}
To investigate connections between the unfolding epidemic and socio-demographic, infrastructural, and environmental factors, we focus on six scalar covariates that can be retrieved from public sources at the resolution of provinces with reasonable data quality (see Table~\ref{supp_tab:scalar_covariates}). 
Each covariate works as a proxy for a potentially relevant factor, namely: aging of the population (\texttt{$\%$ Over 65}); quality of distributed primary health care (\texttt{Adults per family doctors}), which has been contrasted to that of centralized, hospital-based care; the potential of hospitals (\texttt{Ave.~beds per hospital}), schools (\texttt{Ave.~students per classroom}) and workplaces (\texttt{Ave.~employees per firm}) to act as contagion hubs; 
and pollution levels (\texttt{PM10}). 
The significance of this kind of proxies has been shown in recent studies related to COVID-19, both focusing on the first wave of the Italian epidemic \cite{boschi2021functional, basellini2021explaining}, and on inter-generational relationships using data from 24 countries \cite{arpino2020no}. 
\begin{table}
    \caption{\textbf{Scalar covariates:} six variables considered as proxies of potentially relevant socio-demographic, infrastructural, and environmental factors.}
    \label{supp_tab:scalar_covariates}
     \begin{tabularx}{\textwidth}{lXl}
    \toprule
    
    {\bf Covariate} & {\bf Description} & {\bf Year and Source} \\
    
    \midrule

    \% \texttt{Over 65}
    & Aging of the population  
    &  2020, ISTAT\\
    
    \texttt{Adults per family doctor}
    & Quality of distributed
    %, 
    primary health care 
    &  2019, Ministry of Health\\
    
    \texttt{Ave.~beds per hospital}
    & Ability of hospitals to act as contagion hubs 
    &  2019, Ministry of Health \\
    
    \texttt{Ave.~students per classroom}  
    & Ability of schools to act as contagion hubs 
    &  2019, Ministry of Education \\
    
    \texttt{Ave.~employees per firm}
    & Ability of work places to act as contagion hubs 
    &  2018, ISTAT \\
   
    \texttt{PM10}  
    & Pollution levels (particulates)
    &  2019, ISTAT \\
    
    \bottomrule

    \end{tabularx}
\end{table}
In particular, the variable \texttt{\% Over 65} is retrieved from ISTAT \citep{istat_over65} at provincial level. 
To compute \texttt{Adults per family doctor}, we divide the population of the province as of January 2019 (from ISTAT) by the number of family doctors in the province \citep{istat_ASC}. The latter is obtained by summing the number of family doctors at the level of ``Aziende Sanitarie Locali'' (Local Health Agencies, ASL) provided by the Ministry of Health for the year 2018. We manually retrieved data missing for some provinces in Lombardia, Molise, Sardegna, and Toscana from the corresponding ASL websites.
To compute \texttt{Ave.~beds per hospital} we use data from the Ministry of Health \citep{hm_hosp}, which provides the number of beds per ward in each hospital in 2019. We first aggregate them over wards belonging to the same hospital, and then average over hospitals in each province. 
To compute \texttt{Ave.~students per classroom} we use data from the Ministry of Education \citep{min_educ}, which provides the number of students in each classroom of each school in the country (public or private, at every level of education), for the academic year 2019/2020. We average them over schools in each province. Missing data for Trento, Bolzano, and Aosta are filled through random forests imputation \citep{stekhoven2012missforest}, with default parameters \texttt{maxiter=10} (maximum number of iterations performed if the stopping criterion is not met) and \texttt{ntree=100} (number of trees grown in each forest).
To compute \texttt{Ave.~employees per firm} we use data from ISTAT \citep{istat_ASC}, which provides the number of employees per firm at the level of municipalities. We average them over firms in each province.
To compute \texttt{PM10} we use data from ISTAT \citep{istat_air} -- which provides the average annual concentrations of PM10 (in $\mu g/m^{3}$) detected by air quality meters distributed over the Italian territory -- and we averaged them over meters located in each province. 

\subsection{Data on local mobility}
\label{subsec:mobility}
We use data on daily differential mobility provided by Google at the resolution of Italian provinces \citep{google_mob}. 
These measure the fractional reduction of mobility with respect to levels registered in the first five weeks of 2020 (January 3 to February 6), and are organized in categories based on mobility aims. 
Specifically, we focus on two categories that capture the local, short-range mobility of individuals, which was allowed in all provinces also during mobility restrictions: 
%the category 
``Grocery \& Pharmacy", which accounts for local trips to grocery and pharmacy stores, and 
%the category
``Workplace", which  accounts for trips to workplaces. 
We obtain mobility curves by smoothing these daily data with the same procedure used for mortality curves (see Section \ref{subsec:data_deaths}).
% \marzia{[Are these mobility curves aligned?]} \jacopo{[I remember that yes]} \lorenzo{We align them at a later stage, and we write it. See Section 3.4 and 3.5}

\section{Results}
\label{sec:results}

\subsection{Inconsistencies in death and case data}
\label{subsec:inconsistencies}
\begin{figure}[!t]
\centering
\includegraphics[width=\linewidth]{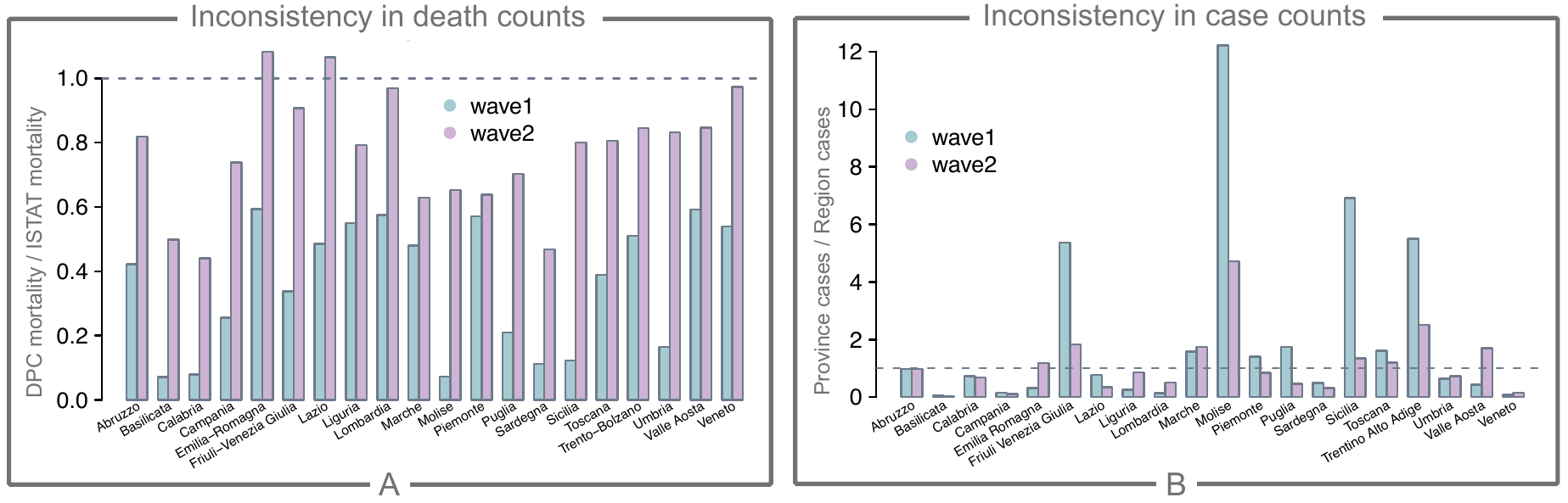}
\caption{
    \textbf{Inconsistencies in death and case counts.}
    Panel \textbf{A} displays, for each region and wave, the ratio between COVID-19 mortality as reported by the DPC and differential mortality based on ISTAT data. During the first wave, all regions show marked inconsistencies between the two sources, suggesting a vast undercounting of deaths in the official records (ratios much lower than $1$). The situation improves during the second wave; even though some regions appear to still under-report deaths (e.g., Basilicata, Calabria, Sardegna). The ratios slightly above $1$ in Emila Romagna and Lazio may reflect reductions in mortality, perhaps indirectly related to COVID-19 (e.g., fewer accidents during lockdowns and curfews), exceeding the epidemic death toll. 
    Panel \textbf{B} displays, for each region and wave, the ratio between DPC case counts obtained by aggregating province-level counts within each region, and DPC case counts provided at the regional level. The inconsistencies (i.e.~departures from $1$) are staggering, especially during the first wave -- when aggregated provincial records exceeded regional records by over 12-fold in Molise, and around 7-fold in Sicilia. As for death counts, the situation improves during the second wave, but inconsistencies remain strong enough to suggest that case count data are not to be relied on.
}
\label{supp_fig:data_quality}
\end{figure}

We compare data from different sources to provide insight into their quality. 
Specifically, the comparison of ISTAT differential mortality and DPC COVID-19 mortality at the resolution of regions %\marzia{and for each wave (i.e., summing up 150-day data for each wave;} see Figure \ref{supp_fig:data_quality}A) 
reveals a vast under-counting in the DPC official reports -- markedly in the first wave, but also in the second \citep{ciminelli2020covid, henry2022variation, modi2021estimating}
(see Figure \ref{supp_fig:data_quality}A; for each region, we contrast totals over the 150 days comprised in each wave). 
Overall, DPC  
%records 
and ISTAT data become more consistent in the second wave, especially in regions such as Emilia Romagna, Lazio, Lombardia, Veneto, and Friuli Venezia Giulia. 
This,
%may, 
by and large, 
%be interpreted as 
suggests an improvement in the official DPC COVID-19 death records -- even though 
%DPC records 
they remain much lower than ISTAT differential mortality in regions such as Basilicata, Calabria, and Sardegna. 
Notably, in Emilia Romagna and Lazio DPC records were slightly higher than ISTAT differential mortality during the second wave. This may be a consequence of the fact that the latter reflects also phenomena indirectly related to COVID-19 -- affecting mortality in both directions; e.g., increases in mortality due to untreated emergencies or untimely treatment of chronic ailments, but also reductions in mortality due to fewer accidents during lockdowns or curfews.

The number of cases reported by the DPC at the two resolutions of provinces and regions harbor massive and concerning inconsistencies. 
As shown in Fig.~\ref{supp_fig:data_quality}B, by aggregating reported province-level cases into regional case counts, we obtain totals that do not match reported region-level cases.
This is strikingly demonstrated by Molise and Sicilia, where the sum of the cases reported in the provinces are, respectively, over $12$ and around $7$ fold higher than the cases reported at regional level. 
Because of the vast discrepancies between provincial and regional case counts, we decided not to consider these data in our analyses.

In short, due to the strong limitations that still exist in the official COVID-19 data, we focus on 
%can consider 
mortality alone (not cases) to characterize the epidemic, and we do so
%need to use 
using a proxy such as differential mortality (not official records) -- 
%that also 
which has its own limitations. 
This may indeed comprise some deaths not related, or indirectly related, to COVID-19 -- but it is still more likely to accurately reflect mortality than the officially released deaths records, and it allows us to perform analyses at the spatial resolution of provinces.

\subsection{Mortality patterns}
\label{subsec:mortality}
Our analyses, consistent with prior studies \citep{borghesi2021lombardy, chirico2021covid}, delineate a dramatic and concentrated first epidemic wave -- with similar timing across a small number of very hard-hit provinces, especially in Lombardia \citep{zirilli2022covid}. In contrast,  the second wave was more ``spread'', with smaller differences across provinces and weaker synchronization of the mortality curves.

\begin{figure}[t]
\includegraphics[width=\linewidth]{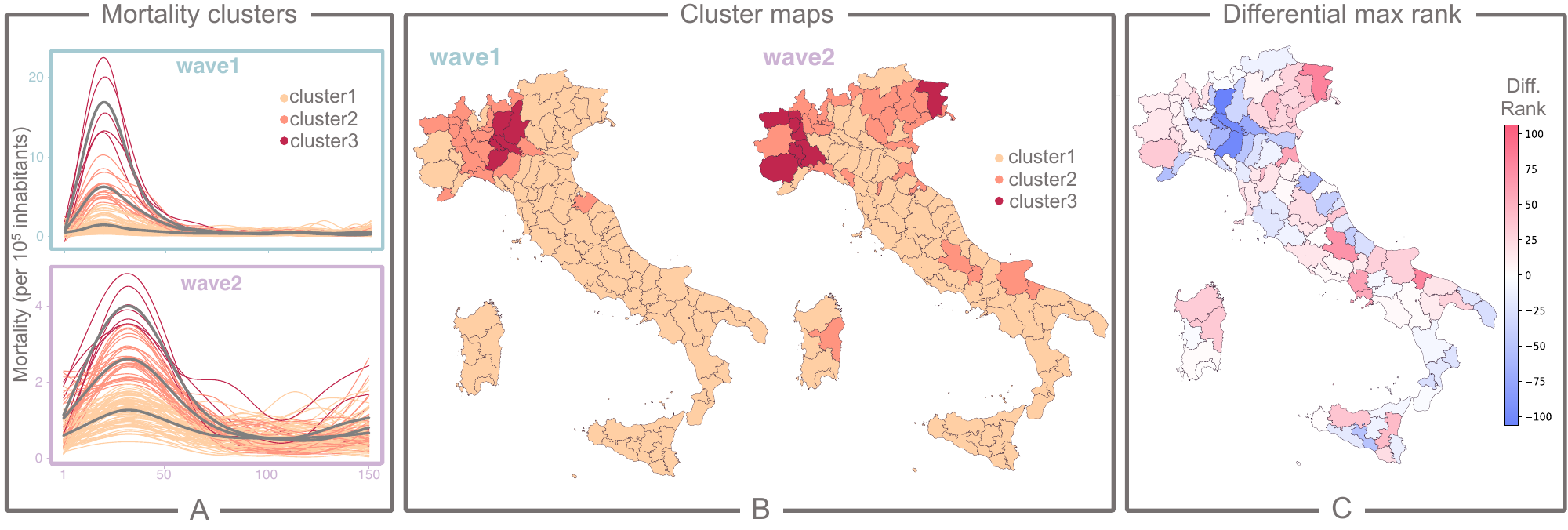}
\caption{
    \textbf{Clustering of mortality curves.}
    Panel \textbf{A} shows aligned province-level mortality curves in the first (top) and second (bottom) waves of the epidemic. For each wave, the curves are color-coded according to a partition in three clusters -- with dark red, orange, and yellow representing decreasing aggressiveness of the epidemic. Thicker gray lines trace the means of the three clusters.
    Panel \textbf{B} shows the three clusters on the Italian map, for the first (left) and second (right) wave. Panel \textbf{C} shows again the Italian map, with provinces color-coded based on the difference in the ranks a province occupied in the first and second wave -- for each wave, we constructed rankings based on the magnitude of the curves' largest peaks. Thus, shades of blue (red) indicate provinces that suffered the first (second) wave more than the second (first).
}
\label{fig:cluters}
\end{figure}

Fig.~\ref{fig:cluters}{A} shows aligned mortality curves during the first and second wave (see also Fig.~\ref{supp_fig:mortaliy_and_shifts} which, in addition, shows the curves before the shifts applied to align their peaks, as well as the shift distributions). 
Notably, the largest mortality peaks were much higher in the first wave (around 20 daily deaths per 100,000 inhabitants) than in the second (around 5 daily deaths per 100,000 inhabitants), highlighting the grim death toll 
%of the first Italian epidemic wave. 
Italy withstood during the former. However, such death toll was very concentrated; many curves remained low, and just a handful had very high peaks. 
In contrast, the mortality peaks of the second wave were less differentiated, indicating an epidemic more evenly spread across the country. 
In terms of timing, the first epidemic wave was markedly more synchronous than the second; this can be appreciated considering the shifts required to align peaks, which are
%in the latter  
larger (a median of 19 vs.~12 days) and more variable in the second wave (see Fig.~\ref{supp_fig:mortaliy_and_shifts}).

We employ clustering techniques to characterize mortality patterns systematically, and further highlight differences in the unfolding of the two epidemic waves. 
We group shifted mortality curves in the first and (separately) the second wave using an agglomerative hierarchical clustering algorithm with $L^2$ (i.e.~Euclidean) distance and complete linkage. 
%More in detail, 
In more detail, the $L^2$ distance between two generic curves $x$ and $v$ in the interval $[0,c]$ is defined as
\begin{equation}
    d(x,v)=\frac{1}{c} \int_0^c \left( x(t)-v(t) \right)^2 dt
    %;
    \ .
\end{equation}
%starting 
Starting from the matrix of pairwise distances, the agglomerative algorithm merges individual curves based on their distance, and clusters of curves based on the maximal distance between their members. 
The resulting dendrograms, one for each wave, are shown in Fig.~\ref{supp_fig:provinces_dendrograms}. These are cut to obtain %k=2$ or 
$k=3$ clusters in each wave, based on the dendrogram structure and the Hartigan index \citep{hartigan1975clustering}.
The province-level aligned mortality curves in Fig.~\ref{fig:cluters}{A} are color-coded based on $k=3$ clusters in each wave, and Fig.~\ref{fig:cluters}{B} shows the same clusters on the Italian map (cluster memberships for all provinces are reported in Fig.~\ref{supp_fig:provinces_clusters}). 
As documented by 
%other 
prior studies performed at the regional level \citep{boschi2021functional}, in the first wave Italy saw the unfolding of two very different
%epidemics
epidemic patterns: a relatively mild one in the majority of the country, and a very strong one affecting the north-western areas. 
%-- which could be further split into two patterns based on severity. 
%Our 
The current analysis, performed at the higher resolution of provinces, affords a more detailed description of this phenomenon: 
while the epidemic was more severe and more spatially concentrated in the first wave than in the second,
%(weaker and more spread around the country in the second) 
both waves harbored two, possibly three different patterns; namely, 
%are characterized by 
a flatter, ``mild'' pattern 
%\marzia{
in the majority of the provinces, a very steep, ``exponential'' one in a small number of provinces, and an intermediate, still ``exponential'' but less extreme pattern in the rest of the country.
%} 
%However, the epidemic was more severe and more spatially concentrated in the first wave, and weaker and more spread around the country in the second.

In more detail, the provinces that withstood the most dramatic losses during the first wave were Bergamo, Brescia, Cremona, Lodi, and Piacenza. 
These provinces (in red in Figs.~\ref{fig:cluters}{A} and \ref{fig:cluters}{B}) are all contiguous, in an area between Lombardia and adjacent Emilia Romagna (the very first cases of COVID-19 in the country were reported in Codogno, a municipality in the province of Lodi). 
Interestingly, in the second wave, all these provinces belong to the cluster with the lightest mortality (in yellow in Figs.~\ref{fig:cluters}{A} and \ref{fig:cluters}{B}). 
This could be due to several factors. 
The inhabitants of these areas may have been more diligent in practicing social distancing rules and adhering to restrictions and safety mandates after the dramatic events of the first wave. 
In addition, the large number of deaths among vulnerable individuals during the first wave may have effectively reduced the number of people at risk in these areas. 
Finally, the large number of cases during the first wave may have induced some degree of herd immunity \citep{randolph2020herd}.
In the second wave, the hardest hit provinces were Alessandria, Aosta, Asti, Biella, Cuneo, Udine, and Vercelli. 
Interestingly, the mortality curves for all these provinces but Vercelli are also characterized by a very large ``shoulder''. 
The change in hardest-hit provinces with respect to the first wave, along with the fact that fewer provinces belonged to the lightest mortality cluster (in the first wave the yellow, orange, and red clusters, from lightest to hardest hit, comprised 86, 16 and 5 provinces, respectively; these counts for the second wave were 72, 28 and 7), confirm that, while less severe in terms of mortality, the epidemic had a broader spread across the country in the second wave. 

We further compare the two waves considering the ranking of the provincial level mortality curves based on their peaks (maxima; from smallest to largest) and the differences between such rankings -- specifically, ranking in the second wave minus ranking in the first. 
These differences are color-coded on the Italian map of Fig.~\ref{fig:cluters}{C}, and show very clearly how many among the provinces most affected in the first wave (highest ranks) were among the least affected in the second (lowest ranks). The province of Bergamo is the clearest example; it ranked 107$^{th}$ in the first wave and 1$^{st}$ in the second. 
In contrast, some provinces in Piemonte were hard hit (high ranks) in both first and second waves -- showing very little difference in ranking.

\subsection{Role of restrictions and socio-demographic, infrastructural and environmental factors}
Next, we analyze potential associations between mortality and the socio-demographic, infrastructural, and environmental factors listed in Table~\ref{supp_tab:scalar_covariates}.
In doing so, we also consider two additional variables calculated from the mortality curves themselves, which we call \textit{area before}, $A_{bef}$, and \textit{area after}, $A_{aft}$. 
For each province, these measure the areas under the mortality curve to the left and the right, respectively, of the date when restrictions were introduced (March 9, 2020, for the first wave and November 4, 2021, for the second). 
Hence, they represent the cumulative mortality at the time of restriction and after restriction, respectively. As these variables are highly right-skewed, we take their logarithms to normalize them.
Intuitively, before the introduction of restrictions the epidemic accelerates, with contagions spreading and, metaphorically, filling a pool of ``potential deaths''. Restrictions, ideally, stop this process -- or rather, in practice, curb it with varying degrees of effectiveness. After their introduction, as the mortality pattern unfolds over time, the pool of potential deaths empties. 
$A_{bef}$ is our attempt at capturing how far the epidemic has accelerated before restrictions hit the breaks on it -- and
%we 
to do so based on mortality data, not cases, because data on the latter are so poor and unreliable. We use
%it 
$A_{bef}$ as a potential predictor in our regression models, along with socio-demographic, infrastructural, and environmental factors. In a way, it can be  thought of as an ``intermediate" predictor, subsuming the effects of a multitude of observed and unobserved factors, and influencing the subsequent unfolding of the mortality pattern. $A_{aft}$ is our way to define a scalar summary of such pattern, capturing its cumulative severity. We use it as a simpler, scalar response -- in alternative to the full mortality curve (functional response).   
$A_{bef}$ is akin in spirit to other proxies used in recent studies to emphasize the importance of timely restriction policies in contrasting the spread of the epidemic. 
For example, using the first wave $R_t$ and case data that are not publicly available, a recent study\cite{cintia2020relationship} showed a strong relationship between the time spent above a given epidemic threshold prior to the introduction of restrictions and the total number of confirmed infections. 
Similarly, another recent study\cite{basellini2021explaining} 
%pinpoint 
pinpointed the onset of the epidemic in an area and showed that the delay in such onset (and thus the shortening of the time between onset and restrictions) is a significant negative predictor of mortality levels.

\begin{figure}[tb]
\centering
\includegraphics[width=1\linewidth]{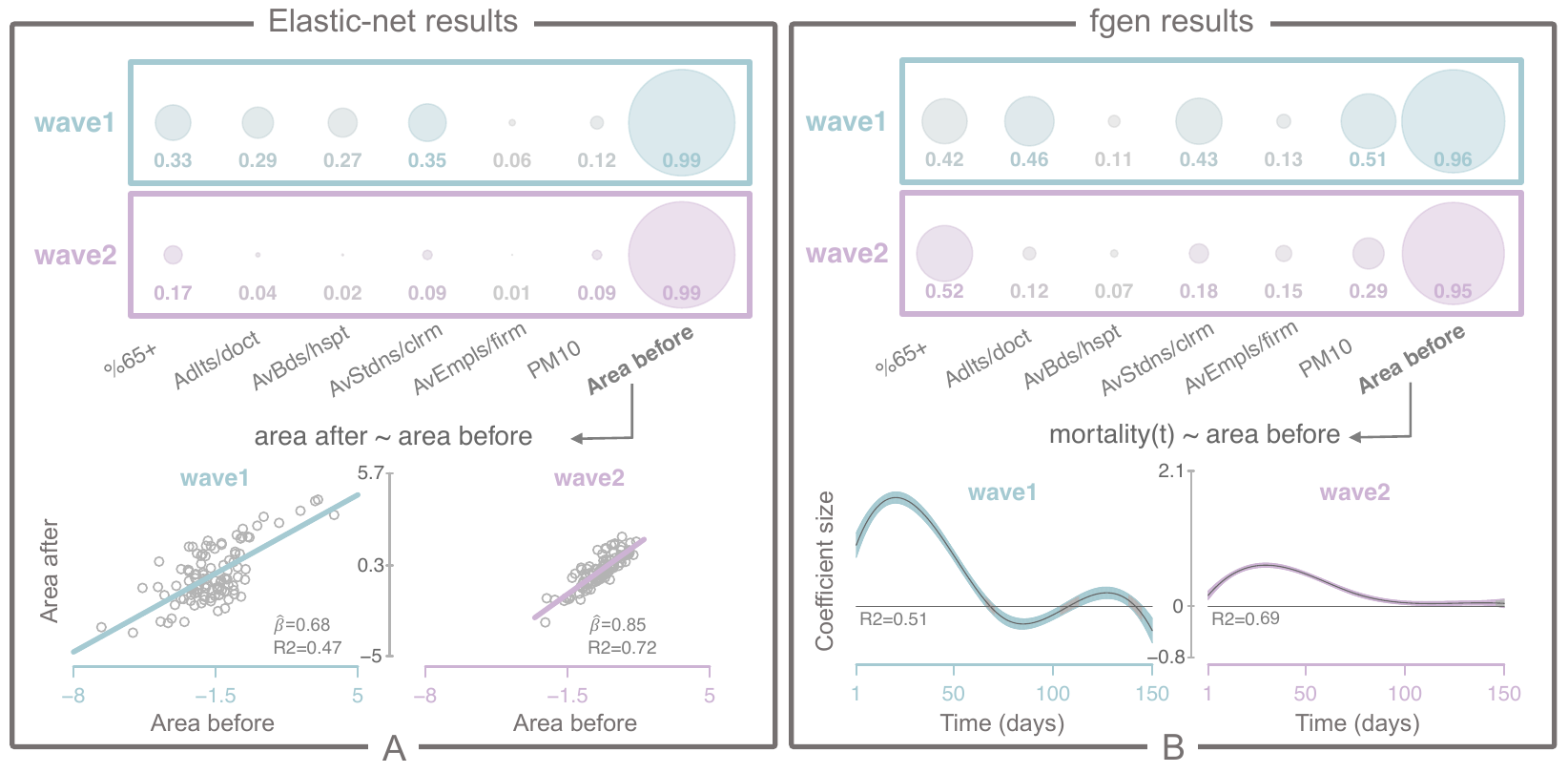}
\caption{
    \textbf{Feature selection results.} 
    Panel \textbf{A}: Elastic net results for $A_{aft}$ (\textit{area after}, scalar response).
    On the top, for each wave and each feature, we show balls with radius proportional to the average $\lambda_{max}$-ratio at which the feature enters the model across 500 replications of the regression on subsets of provinces ($\lambda_{max}$ is the smallest penalty value associated with $0$ selected features; the larger the $\lambda_{max}$-ratio, the earlier a feature is selected and the greater its relevance).
    On the bottom, for each wave, we show fitted line, coefficient estimate $\hat{\beta}$ and $R^2$ for the marginal regression of $A_{aft}$ on $A_{bef}$ (\textit{area before}) -- the most relevant feature in both waves -- run an all 107 provinces (marginal regression results for other predictors are provided in Fig.~\ref{supp_fig:scalar_betas_and_pc1}). 
    Panel \textbf{B}: Functional elastic net (\texttt{fgen}) results for mortality curves, in a format similar to Panel \textbf{A}. 
    For the marginal regressions of mortality curves on $A_{bef}$, bottom plots show coefficient curve estimates $\hat \beta(t)$ as black solid lines, with bands built adding and subtracting $1.96 \times$ point-wise standard errors (marginal regression results for other predictors are provided in Fig~\ref{supp_fig:function_on_scalar_betas}).
} 
\label{fig:feature_selection}
\end{figure}

To evaluate the relative importance of the factors considered in our study, we perform feature selection in two separate regression models. 
%FC: let's not write the models down explicitly.
%\marzia{[Shall we explicitly write the models, in addition to the minimization problems?]} \lorenzo{[I think it may be a sort of repetition, but if you believe that it is clearer, I can easily add them. Let me know what you prefer!]} \marzia{[It is indeed a sort of repetition, I'm just unsure the functional model is clear. Let's see what the others think!]} \tobia{[TB: From my point of view, the model is pretty clear. But maybe I am a bit biased.]}
In the first, we use $A_{aft}$ as scalar response, and implement an elastic net penalized fit \citep{zou2005regularization} with the \texttt{R} package \texttt{glmnet} \citep{friedman2021package}. 
In symbols, the corresponding minimization problem is 
\begin{equation}
     \min_\beta \left( \frac{1}{2} \norm{y - X\beta}_2^2 + \lambda_1 \norm{\beta}_1 + \frac{\lambda_2}{2} \norm{\beta}_2^2 \right),
\end{equation}
where $n$ and $p$ are the number of observations and features, respectively, $y \in \mathbb{R}^n$ is the response vector, $X \in \mathbb{R}^{n\times p}$ is the (standardized) design matrix, and $\beta \in \mathbb{R}^p$ is the coefficient vector. 
We set $\lambda_2 = 0.6 \lambda_1$, and we run the fit for different values of $\lambda_1$, starting from $\lambda_{max}=\lVert X^t y\rVert_\infty$, which selects $0$ features, and gradually decreasing it. 
To capture the relevance of each feature, we track the $\lambda_{max}$-ratio for which it enters the model, defined as $\lambda_{max}\text{-ratio}=\lambda_j/\lambda_{max}$, where $\lambda_j\in(0, \lambda_{max})$ is the largest penalty weight such that feature $j$ is included in the model; a higher ratio (earlier inclusion in the model) corresponds to higher relevance. 
Finally, to assess stability of the results, we repeat the procedure $500$ times, each time using a random subset of $\Tilde{n}$ provinces, with $\Tilde{n}$ itself randomly selected between $90$ and $107$. By removing a random number of units randomly selected without replacement, we aim to mitigate the effect of potential outliers on our findings.
In the second regression model, we use the full mortality curve as functional response, and employ \texttt{fgen} \citep{boschi2021highly}, which is an extension of the elastic net to settings where the response is functional. 
In this case, the minimization problem is 
\begin{equation}
     \min_{\beta_1(t),\dots,\beta_p(t)} \left( \frac{1}{2} \Big\lVert{y(t) - \sum_{j=1}^p X_j\beta_j(t)}\Big\rVert_{{L}^2}^2 + \lambda_1 \sum_{j=1}^p \norm{\beta_j(t)}_{{L}^2} + \frac{\lambda_2}{2} \sum_{j=1}^p \norm{\beta_j(t)}_{{L}^2}^2 \right),
\end{equation}
where $y(t) = [y_1(t),\dots,y_n(t)]$ is the collection of $n$ response curves, $X$ is again the (standardized) scalar design matrix, $\beta(t) = [\beta_1(t),\dots,\beta_p(t)]$ is the collection of $p$ coefficient curves, and $\lVert f\rVert_{{L}^2} = ( \sum_{i=1}^n \lVert f_i\rVert_{{L}^2}^2 )^{1/2}$ is the ${L}^2$-norm for a generic function $f$.
We handle $\lambda_1$ and $\lambda_2$ as in the scalar regression, track again the $\lambda_{max}$-ratio for each feature, and perform the same stability assessment repeating the procedure $500$ times on random subsets of provinces. 
Results for both regressions, each fitted separately on first and second wave data, are summarized in the top graphics of Fig.~\ref{fig:feature_selection}{A} (elastic net for $A_{aft}$) and Fig.~\ref{fig:feature_selection}{B} (\texttt{fgen} functional elastic net for the mortality curves). 
Specifically, for each regression and wave, we display the features' average $\lambda_{max}$-ratios across 500 runs on subsets of provinces. 
The 
%``intermediate'' 
% \lorenzo{[This has to do with causal paths (if I remember well). It mediates the effect on the response of some other (primary) variables. As we do not investigate this causal path in any real way, I do not think it is a necessary adjective. I just comment it.]}
predictor $A_{bef}$ emerges as the most relevant by far in both regressions and in both waves, with the largest average $\lambda_{max}$-ratios. 
Table \ref{tab:elastic_net} and Figure \ref{supp_fig:joint_fgen} report results of the elastic net and \texttt{fgen} feature selections obtained on the full data set comprising all the 107 provinces and with the parameter $\lambda_1$ chosen by 5-fold cross-validation.

As a parallel exercise, for each response and wave, we run marginal regressions on every predictor separately -- using the full set of 107 provinces. 
Results concerning $A_{bef}$ are shown in the bottom graphics of Figs.~\ref{fig:feature_selection}{A} and \ref{fig:feature_selection}{B}. 
Specifically, we display scatter plots with fitted lines, coefficient estimates and $R^2$s for the marginal regressions of $A_{aft}$ on $A_{bef}$ in 
%the 
first and second wave,
%waves, 
and coefficient curve estimates and $R^2$s for the marginal functional
%regression 
regressions of mortality curves on $A_{bef}$, again in first and second wave. 
Marginal results are consistent with the elastic net feature selection, confirming strong positive (i.e.~aggravating) statistical effects of $A_{bef}$ on both waves of the epidemic. In particular, $A_{bef}$ has the largest marginal $R^2$'s among all predictors considered; results concerning all other predictors are provided in Fig.~\ref{supp_fig:scalar_betas_and_pc1} ($A_{aft}$) and in Fig~\ref{supp_fig:function_on_scalar_betas} (mortality curves).

The primacy of $A_{bef}$ in our analyses provides further evidence that the timing of restrictions may play a crucial role in modulating the epidemic. 
Comparing marginal and joint results between the two waves, its effect appears to be larger in magnitude in the second wave for the scalar response, and in the first wave for the functional response. 
The $R^2$'s are always larger in the second wave though, likely due to the fact that then both $A_{aft}$ and the mortality curves exhibited less variability.
%\francesca{across provinces [CORRECT?]} %\lorenzo{[Curves in the second wave are (vertically) smaller than the ones in the first wave. Thus, in principle, in the second wave there is less variability both across and within provinces. Se sono schiacciate tutte quante verso il basso, anche se vibrano/oscillano relativamente di più rispetto a quelle della prima ondata, sono comunque meno variabili in termini assoluti]}.
%FC: ok, lasciamo genereicamente  ``variability'' allora.
%
%%%%
%\textcolor{red}{[AGAIN, WE NEED TO EXPLAIN/CLARIFY SOMETHING HERE; WHY ARE WE REPORTING THE MARGINAL ESTIMATED COEFFICIENTS INSTEAD OF THE JOINT ONES?]}. the exercise is the following: on one hand, rank feature importance with Elastic Net/fgen; on the other, show the marginal magnitude of each feature.
%%%%
%%%INTERNAL NOTE%%%
%FOR BOTH SCALAR AND FUNCTIONAL REGRESSIONS, WE NEED TO EXPLAIN/CLARIFY WHY WE ARE RUNNING A FEATURE SELECTION AND THEN REPORTING MARGINAL RESULTS INSTEAD OF JOINT ONES; I UNDERSTAND THIS IS THE EXERCISE WE PERFORMED, BUT A REVIEWER WILL WANDER
%%%%%%%%%%%
%%%%
%
In addition to being generally weaker, the roles of socio-demographic, infrastructural, and environmental factors appear to vary depending on the response, as well as the wave.
For instance, considering mortality curves, adults per family doctor and average students per classroom exhibit a strong signal in the first wave, along with \textit{PM10} -- confirming previous findings \cite{boschi2021functional}. 
However, the relevance of these predictors decreases in the second wave, where the percentage of inhabitants over 65 seems to play a stronger role.
These differences may result in part from the existence of collinearity among features (see Figs.~\ref{supp_fig:heatmap} and \ref{supp_fig:scalar_coll}), with the %``intermediate'' 
variable $A_{bef}$ possibly subsuming a number of other effects, but they may also reflect actual changes that occurred between the two waves. 
For instance, average students per classroom (as measured in 2019; see Table~\ref{supp_tab:scalar_covariates}) may indeed have been a good proxy for the ability of schools to act as contagion hubs during the first wave, but may have become a much less meaningful predictor during the second wave -- when classroom sizes were reduced, many activities were held online, and even when attending school in person students had to adhere to masking and social distancing protocols.

Finally, we compute \textit{pc1} -- the first principal component of the socio-demographic, infrastructural, and environmental factors in Table~\ref{supp_tab:scalar_covariates}, 
%($A_{bef}$ is not included in this computation), 
which can be used as an overall summary to control for their effects ($A_{bef}$ is not included in this computation). 
\textit{pc1}, which explains $65.6\%$ of the overall variance, is mainly 
%captures the variability of 
driven by adults per family doctor and average beds per hospital (see loadings in Fig.~\ref{supp_fig:scalar_betas_and_pc1}{B}), and
%it 
presents only a very mild correlation with $A_{bef}$ (see Fig.~\ref{supp_fig:response_over_areabefore_pc1}{C}).
We therefore further gauge the role of $A_{bef}$ by running both the scalar regressions (with response $A_{aft}$) and the functional regressions (with response the mortality curves) 
%with 
using $A_{bef}$ and \textit{pc1} as joint predictors, and computing their \emph{partial} $R^2$s. Results from these joint fits (see Fig.~\ref{supp_fig:response_over_areabefore_pc1}) support our previous statements, highlighting once again the importance of $A_{bef}$.
%-$R^2$s.
%(
As a reminder, the partial $R^2$ 
%for 
of a predictor is defined as $(R^2 - R^2_{red})/(1 -R^2_{red})$, where $R^2$ is the coefficient of determination of the complete model, and $R^2_{red}$ that of the model comprising all predictors but the one being evaluated.
%).
%Noticeably, 
%Results for these joint fits (see Fig.~\ref{supp_fig:response_over_areabefore_pc1}) support our previous statements, highlighting once again the importance of $A_{bef}$.

\subsection{Mobility patterns}
\begin{figure}[!t]
\centering
\includegraphics[width=1\linewidth]{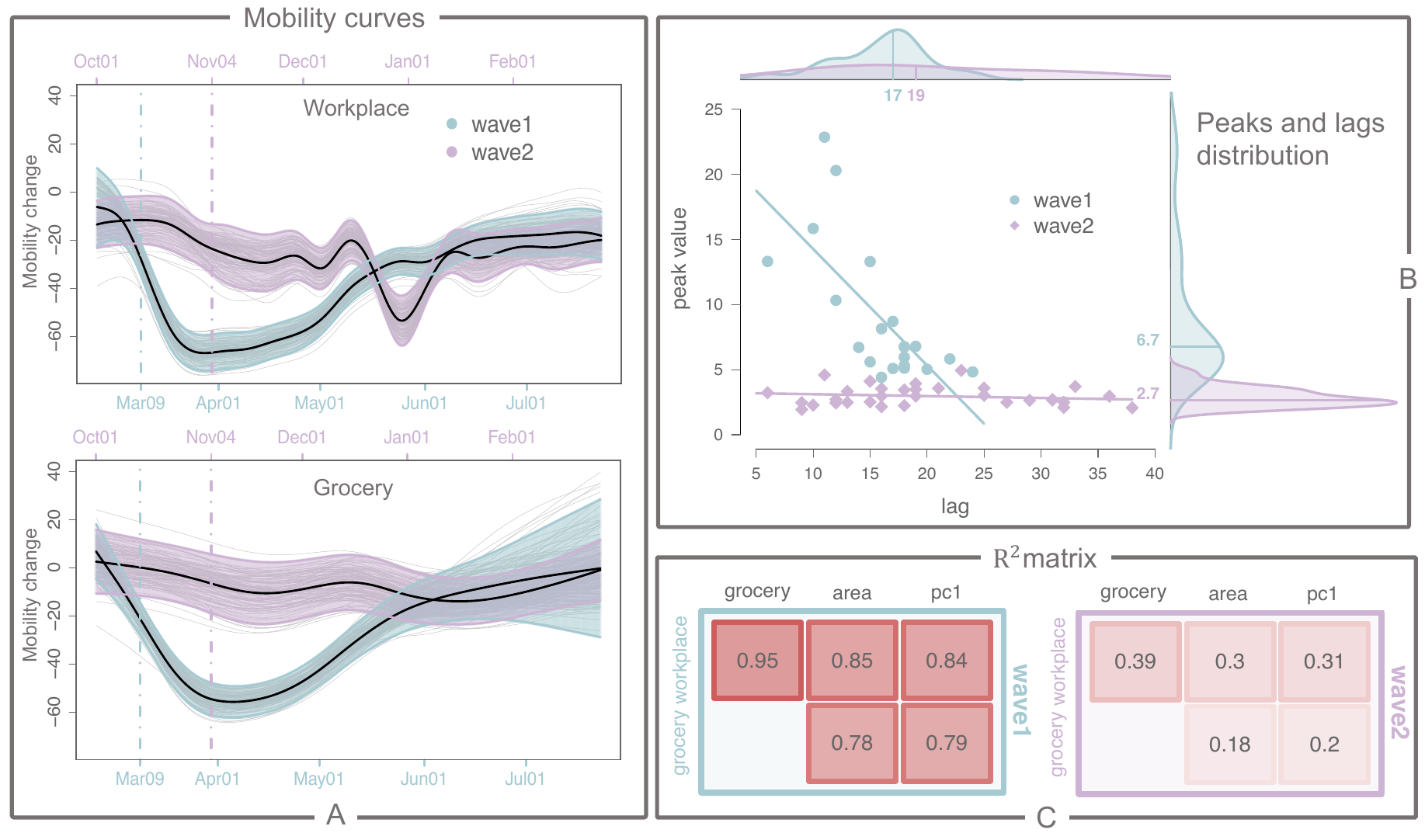}
\caption{
    \textbf{Mobility and lags.} 
    Panel {\textbf{A}}: ``Grocery \& Pharmacy'' and ``Workplace'' mobility curves during first and second wave. Solid black lines are functional means, with bands built by adding and subtracting $1.96 \times$ point-wise standard deviations. Vertical lines mark dates when mobility restrictions were introduced (March 9, 2020, and November 4, 2020, for first and second wave, respectively).
    Panel \textbf{B}: scatterplot and marginal distributions of peak values and lags for provinces with a discernible epidemic (hard- and medium-hit clusters 2 and 3; see Section~\ref{subsec:mortality}) in first and second wave. Lags are defined as the number of days between the mobility restriction date and the mortality peak for each province. Least-squares regression lines on the scatterplot have negative estimated slopes of -0.89 (significant) for the first wave and -0.01 (non-significant) for the second. Vertical lines on the marginal distributions mark medians.
    Panel \textbf{C}: $R^2$ values, obtained by regressing each functional variable in the rows against each functional and scalar variable in the columns, in first and second wave. These are used to gauge predictor collinearities prior to modeling mobility curves. Background color intensity in each block is proportional to the $R^2$.
    %%% INTERNAL COMMENT %%%
    %FOR SOME RELATIONSHIPS IT MAY BE HARD TO GIVE A ``SIGN'' BECAUSE THAT'S NOT UNIFORM ACROSS THE TIME DOMAIN, BUT WE MAY BE ASKED TO COMMENT ON SIGNS!
    %%%%%%%%%%%%%%%%%
}
\label{fig:mobility}
\end{figure}

Several studies \citep{nouvellet2021reduction, cuellar2022assessing, rahman2022associations} found evidence that local, short-range mobility worked as a key modulator of COVID-19 spread. 
We therefore produce province-level mobility curves based on Google's ``Grocery \& Pharmacy" and ``Workplace" indexes -- which express percentage changes in mobility linked to these categories with respect to a pre-pandemic reference (the first five weeks of 2020; see Section~\ref{subsec:mobility}). 
Before analyzing how mortality relates to local mobility (see Section~\ref{subsec:mob_mort} below), we compare patterns in the latter between the two waves, highlighting some marked differences.
During the first wave, local mobility showed a dramatic reduction with respect to reference values, suggesting that the restrictions implemented by the government were generally effective in changing people's behaviors.
In more detail, the mobility curves in Fig.~\ref{fig:mobility}A show the ``Grocery \& Pharmacy" index reaching levels between $-50\%$ and $-60\%$ (depending on the province) during the first wave lockdown. 
This reduction occurred even though citizens were allowed to leave their homes for these necessities.
%also during its most restrictive phases. 
In the same period, and in particular after the halting of all non-essential production activities on March 23, 2020, the ``Workplace" index reached even lower levels, dipping to values between $-60\%$ and $-80\%$.
%(between approximately $-0.80$ and $-0.60$ depending on the province) principally after the halting of all non-essential production, industries, and businesses.
%
In contrast, during the second wave, while local mobility still contracted with respect to the reference, reductions were less dramatic. 
%-- likely due to the milder, adaptive restrictions imposed by the color-coded system and to the fact that citizens took fuller advantage of whatever mobility was allowed to them.
The ``Grocery \& Pharmacy" curves reached minima between $0\%$ and $-20\%$.
Those of the `Workplace" index again dipped deeper, between $-40\%$ and $-60\%$, but only around Christmas and New Year -- when the entire country was red-coded to try and take advantage of the holidays to curb the second epidemic wave.
%for both the mobility curves, values have always been $0$ or below, suggesting a general decrease in mobility compared with the one recorded during the first five weeks of January 2020, i.e.~before the start of the epidemic, meaning that the government restrictions were generally working. 
%
Also strikingly, the variance across curves was very low during the first wave, and much higher during the second.

In summary, mobility decreased dramatically and uniformly across the country during the first wave lockdown, and only moderately, with more variability from province to province
%, 
-- likely as a consequence of the adaptive restrictions imposed by the color-coded system -- during the second wave. 
Notably, these patterns did not just 
reflect differences in the severity and uniformity of the restrictions themselves, but also differences in citizens' attitudes. 
While during the first wave people limited also activities that were permitted by the lockdown rules, during the second wave they took fuller advantage of whatever mobility was allowed to them.

\subsection{Association between mobility and mortality}
\label{subsec:mob_mort}
We now turn to the analysis of the association between mobility and mortality curves. 
To begin, we attempt to evaluate the delay between changes in the former and their potential impact on the latter. %Fig.~\ref{fig:mobility}{B} shows the scatterplot between lags and peaks, together with their marginal distributions.
For each province in each wave, we compute a {\it lag}, defined as the number of days elapsed between the date in which mobility restrictions were introduced and the date in which the province experienced its mortality peak. 
Restricting attention to the provinces which did experience a discernible epidemic in each wave (those in the hard- and medium-hit clusters 2 and 3; see Section~\ref{subsec:mortality}), Fig.~\ref{fig:mobility}{B} shows a scatterplot 
of peak values against lags, together with their marginal distributions, for the two waves.
For these provinces, lags are much more spread in the second wave than in the first, but their medians are remarkably similar (19 and 17 days, respectively).
%suggesting a consistent timing pattern 
%%in the timing of each epidemic wave.
%%%%%INTERNAL COMMENT%%%%%
%[THE LAGS IN THE SECOND WAVE HAVE SUCH A BROAD AND FLAT DISTRIBUTION THAT THE MEDIAN MAY NOT MEAN MUCH? MAYBE LET'S NOT COMMENT ON THIS].
In contrast, peak values are higher and much more spread in the first wave than in the second; in particular, the first wave peak distribution has a long right tail.
Notably, there appears to be a distinct negative relationship between lag and peak size in the first wave, which provides yet more evidence for the importance of restrictions timing. 
In the provinces most severely affected by the epidemic, restrictions came too late to curb mortality; these provinces reached their very high mortality peaks shortly after the implementation of the national lockdown. 
In contrast, provinces that, while affected, were spared the worst impacts were those where restrictions came earlier during the epidemic progression; they had smaller peaks occurring at a later stage.  
Intriguingly, an analogous negative relationship between lags and peaks cannot be traced in the second wave, when even in the worst-hit provinces peak values remained relatively low. 
%\textcolor{red}{[SO DATA FROM THE SECOND WAVE DOES NOT SUPPORT THE RESTRICTION TIMING ARGUMENT? THE FACT THAT THERE IS LESS VARIABILITY IN PEAK VALUES MAY PARTIALLY EXPLAIN THIS, BUT WE NEED TO UNDERSTAND THIS BETTER. IF WE FIT A LINE DO WE DETECT A SMALL BUT SIGNIFICANT NEGATIVE SLOPE FOR THE SECOND WAVE?]}. 
%\textcolor{green}{[The fitted coefficient is very slightly negative, but not significant. I guess that less variability in peak values is the main point here.]}

Next, we attempt to evaluate the impact of mobility on mortality developing appropriate regression models. 
We consider as potential predictors 
%for 
of mortality curves both ``Grocery \& Pharmacy'' and ``Workplace'' mobility curves, as well as two scalar variables from our prior analyses; $A_{bef}$ (which summarizes the progression of mortality up to the introduction of restrictions), and {\it pc1} (which summarizes the factors listed in Table~\ref{supp_tab:scalar_covariates}). 
To gauge collinearities, we  fit functional regressions of ``Grocery \& Pharmacy'' and ``Workplace'' on each other, and on each of the scalar variables. 
Fig.~\ref{fig:mobility}{C} shows the resulting $R^2$s which, together with the correlations reported in Figs.~\ref{supp_fig:scalar_coll} and \ref{supp_fig:response_over_areabefore_pc1}{C}, provide an overall picture of the dependence structure among explanatory variables. In the first wave, predictors show very strong linear associations. 
In particular, ``Grocery \& Pharmacy'' and ``Workplace'' mobility exhibit an $R^2$ of $0.95$, confirming the effectiveness of the first lockdown.
Both mobility indices are also highly correlated with $A_{bef}$ (the $R^2$ is $0.78$ for ``Grocery \& Pharmacy'' and $0.85$ for ``Workplace''). 
A possible explanation is that, while mobility plummeted everywhere, whatever variation existed in such reduction across the country saw places hardest hit by the epidemic up to the lockdown date contract mobility the most. %\textcolor{red}{[IS THIS A PLAUSIBLE EXPLANATION? IS THIS THE OBSERVED ASSOCIATION SIGN?]}. 
%\textcolor{green}{[It is indeed a possible explanation. I confirm that the association sign is positive, so a decrease in mobility is associated with a decrease in area before.]}
Because of these strong associations, to avoid variance inflation phenomena, we consider a very simple model for the first wave (Model A) -- regressing mortality curves only on one type of mobility curves, say ``Workplace", and \textit{pc1} as a control variable. %\textcolor{red}{ [WARNING! THE $R^2$ BETWEEN WORKPLACE MOBILITY AND $pc1$ is $0.84$!]}. \textcolor{green}{[true... but it is also true that adding and removing pc1 does not change anything in terms of coefficient estimation]}
In the second wave, predictors show much milder 
%%% INTERNAL COMMENT%%%
%positive?
%%%%%%%%%%%%%%%%%%%
linear associations. 
The more nuanced and differentiated color-coded restrictions allowed ``Grocery \& Pharmacy'' and ``Workplace'' mobility to vary more across the country and to capture different signals, with an $R^2$ decreasing to $0.39$. 
Also the $R^2$s with $A_{bef}$ decrease (to $0.18$ for ``Grocery \& Pharmacy'' and $0.30$ for ``Workplace'').
We therefore consider both the simple model used for the first wave (Model A), and a more complex model (Model B) -- regressing mortality curves jointly on both types of mobility curves, $A_{bef}$ and again \textit{pc1} as a control.

In both Model A and Model B we also introduce a binary predictor $d$ to separate provinces with discernible epidemics (hard- and medium-hit clusters in Section~\ref{subsec:mortality}; $d=1$, Group 2) from the others (mild-cluster; $d=0$, Group 1). 
Finally, we formulate both models as ``lagged'' concurrent function-on-function regressions \citep{ramsay2005, horvath2012inference}. 
The functional response (mortality) at time $t$ is thus assumed to depend only on a functional predictor (mobility) at time $t-\ell$, where $\ell$ (the lag) is an additional model parameter. 
In symbols, the general formulation we utilize is 
\begin{equation}
    \label{eq:fun_reg}
    y(t) = \alpha(t) + \alpha_d(t)d + \sum_{m=1}^M \Big( \beta_{m}(t - \ell) X_{m}(t-\ell) + d \beta_{m,d}(t-\ell) X_{m}(t-\ell) \Big)
    + \sum_{j=1}^J \Big( \beta_{j}(t)X_{j} +  d\beta_{j,d}(t) X_{j} \Big) + \epsilon(t)\,, 
\end{equation}
where, as before, $y(t) = [y_1(t),\dots,y_n(t)]$ is the collection of $n$ shifted mortality curves, $n=107$, and $\epsilon(t)$ are i.i.d.~Gaussian errors. 
$M$ is the number of functional predictors ($M=1$ in Model A and $M=2$ in Model B), all introduced with the same lag $\ell$, and $J$ is the number of scalar covariates ($J=1$ in Model A and $J=2$ in Model B). 
Finally, $\alpha(t)$, $\beta_{m}(t - \ell)$, and $\beta_{j}(t - \ell)$ are functional intercept and coefficient curves for provinces where the epidemic was mild ($d=0$) while $\alpha(t) + \alpha_{d}(t)$, $\beta_{m}(t - \ell) + \beta_{m, d}(t - \ell)$, and $\beta_{j}(t) + \beta_{j, d}(t)$ are the corresponding curves for provinces where the epidemic was hard/intermediate ($d=1$). 
Before fitting our models, we shift mobility curves by applying, for each province, the same shift used to align mortality curves (Fig.~\ref{supp_fig:mobility_shifted} shows the mobility curves after shifting).
Moreover, based on the distributions of lags reported in Fig.~\ref{fig:mobility} and on previous findings \citep{carteni2021role}, we fit the models with 10 different values of $\ell$ ranging from 15 to 24 days
%15 (included) to 25 (excluded) days
%\tobia{
-- note that this range includes the medians of the lag distributions 
%of 
for both waves.
Estimating coefficient curves at different lags addresses potential stability pitfalls in modeling functional objects and provides a more robust understanding of the relationship between the investigated variables \citep{yu2020veridical}.
Indeed, testing various lags allows us to determine whether the estimated functional coefficients are heavily influenced by the specific lag, or whether they exhibit a consistent effect that is not contingent on the specific value of $\ell$ within the range considered.
%} 
To fit functional regressions we use the \texttt{R} package \texttt{refund} \citep{goldsmith2016refund}, which estimates 
%the 
functional coefficients
%as well as 
and their standard errors.

\begin{figure}[t]
\centering
\includegraphics[width=1\linewidth]{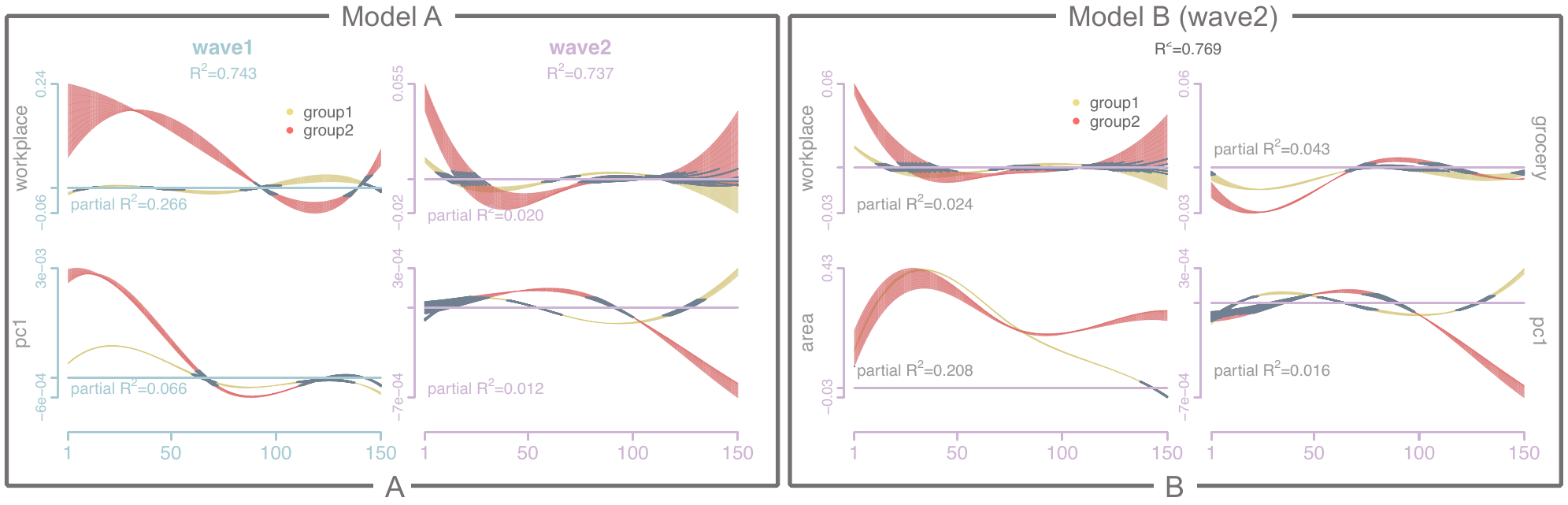}
\caption{
    \textbf{Lagged concurrent functional regression results.} 
    Panel \textbf{A}: for first (left) and second (right) wave, estimated coefficient curves for the regression of mortality on ``Workplace" mobility and \emph{pc1}, with an additional binary predictor separating provinces with mild ($d=0$, Group 1) vs.~hard/intermediate ($d=1$, Group 2) epidemic courses. The plots show ``beams'' comprising $10$ curves, one for each of the lags considered in the model ($\ell=15,\dots,24$). Gray portions in a curve correspond to time intervals where it did not significantly depart from $0$ (the confidence interval obtained adding and subtracting to the estimate $1.96 \times$ pointwise standard errors contained $0$). For each regression we report the average total and partial $R^2$'s over the 10 fits with different lags.
    Panel \textbf{B}: in a format similar to {\bf A}, but for the second wave only, results for the regression of mortality on ``Workplace'' mobility, ``Grocery \& Pharmacy'' mobility, $A_{bef}$ and $pc1$, again including the binary predictor separating provinces with mild vs.~hard/intermediate epidemic courses. 
}
\label{fig:fof_simpler}
\end{figure} 

Fig.~\ref{fig:fof_simpler}{A} summarizes results for Model A fitted (separately) on both waves. 
As expected, coefficient curve estimates are generally larger in magnitude for the first vs.~the second wave, and for hard/intermediate hit vs.~mildly hit provinces. In both waves, ``Workplace'' mobility is the leading predictor variable (based on the partial-$R^2$). 
Its effect is positive along most of the domain during the first wave, but only at the beginning of the second wave. 
Notably, coefficient curve estimates are fairly consistent in shape and magnitude across lags in the 15-to-24 day range, suggesting stability of our findings. 
Robustness to varying model specifications is also confirmed using ``Workplace'' mobility and the group dummy $d$ as predictors (without $pc1$ as control; see Fig~\ref{supp_fig:function_on_function_models}{B}), or replacing ``Workplace'' with ``Grocery \& Pharmacy'' mobility (with $pc1$ and $d$; see Fig.~\ref{supp_fig:function_on_function_models}{A}, or without $pc1$; see Fig~\ref{supp_fig:function_on_function_models}{B}). 
Interestingly though, coefficient curve estimates for ``Grocery \& Pharmacy" 
are slightly negative in the first half of the second wave. 
A mild (and temporary) negative association between this type of mobility and mortality may in fact make sense; there was not a hard policy restriction against these first necessity-type of movements, and the public may have engaged in them more liberally in provinces with lower mortality and a perception of (relative) safety.
Finally, Fig.~\ref{fig:fof_simpler}{B} summarizes results for Model B fitted on the second wave. 
Coefficient curve estimates for both mobility predictors (``Grocery \& Pharmacy'' and ``Workplace''), which are here considered jointly, are consistent with those obtained from fits of Model A. 
However, the predictor with the largest partial $R^2$ here is $A_{bef}$, which shows a strong positive association with mortality -- providing yet more evidence for the critical role of restriction timing. 
Interestingly, the effect size of $A_{bef}$ is similar in hard/intermediate and mildly hit provinces during the first half of the period covered by the second wave. 
This may be due to the definition of $A_{bef}$, which is the integral of the curve in the first part of the domain (before restrictions take place). 
As in Model A, coefficient curve estimates are consistent in shape and magnitude across lags in the 15-to-24 day range, %\tobia{identifying a
indicating a strong and stable effect. Furthermore, 
%the 
total and partial $R^2$s are comparable across 
%the 10 fits with different 
lags. Thus, 
%Both  
both the explanatory power of the model and the relationships between
%the 
response 
%variable 
and 
%the 
predictors are 
%not affected by the specific 
stable across the range of lags considered,
%suggesting stability and
lending reliability 
%of 
to our findings.
%}

%\textcolor{red}{[A REFEREE MAY WANDER WHY WE DIDN'T PICK ONE LAG (THE BEST?) AND NOT EASILY LATCH ONTO THE STABILITY NOTION WE ARE TRYING TO CONVEY. CAN WE SAY SOMETHING MORE? WHAT DOES IT SUGGEST, IN INTERPRETATION TERMS, THAT OUR CURVE ESTIMATES DON'T CHANGE MUCH AS WE CHANGE THE LAG IN THE CONCURRENT REGRESSION FITS? WHERE FITS WILL ALL LAGS PRETTY MUCH EQUIVALENT ALSO IN TERMS OF $R^2$, OR WHERE SOME BETTER?]} \tobia{TB: I tried to address this comment adding some text in blue. Let me know if we should expand further.}

\section{Discussion}
\label{sec:discussion}
Notwithstanding the limited availability and at times poor quality of publicly available data, we were able to leverage techniques from the domain of functional data analysis to characterize and compare the initial two waves of the COVID-19 pandemic in Italy. 
Our analysis is conducted at the provincial level;
this finer resolution compared to prior studies by our group and others \citep{boschi2021functional, azzolina2022regional, della2020network, guzzi2020spatio} allowed us to shed additional light on critical associations and relationships -- avoiding the loss of signal caused by spatial aggregations and enabling the use of more sophisticated statistical models. 
The two waves span a complete year, from February 2020 to February 2021, and both occurred before the beginning of the vaccination campaign. 
Therefore, the patterns and associations we observed were not influenced by vaccine availability and the progressive increase of vaccine-related immunity in the country.

Using hierarchical clustering, we identified three clusters in both waves of the pandemic; one representing a mild mortality pattern, and two representing an intermediate strength and a harsh exponential pattern, respectively. 
Perhaps the most interesting observation obtained by contrasting clustering results for the two waves is that some of the provinces most impacted during the first wave (e.g., Bergamo, Como, and Lodi in Lombardia) were among the least impacted during the second. 
This could have been due to a reduced number of vulnerable individuals after the deaths in the first wave, to behavioral adaptations that led the population in these areas to adhere more strictly to recommendations on, e.g., social distancing and mask use, and possibly to a degree of herd immunity. 

Next, we highlighted the importance of a timely introduction of restrictions. 
Given the poor quality of data on cases, we created a variable, $A_{bef}$, that quantifies the escalation of the epidemic prior to the introduction of restrictions based on mortality data alone. 
Using a number of regression analyses and techniques we found $A_{bef}$ to be a very strong predictor of mortality in both waves -- in fact a stronger and more consistent predictor than the proxies for socio-demographic, infrastructural, and environmental factors at our disposal. 
Potentially suboptimal proxies, collinearity, and possible confounding effects require that we interpret these results with care. 
But they do suggest that, by ``hitting the breaks'' on the exponential dynamics of mortality, restrictions play a pivotal role in curbing the spread of the epidemic and mitigating its impact.

Finally, we explored the relationships between mortality and mobility patterns. 
During the first wave lockdown, mobility exhibited very pronounced and similar contractions across the country. 
Following the introduction of the second wave color-coded restriction system, mobility contractions were milder and more geographically differentiated. 
Despite these differences, through additional functional regression exercises we were able to identify significant and robust positive associations between mobility and mortality in both waves.

This study, alongside our previous work\cite{boschi2021functional}, demonstrates the potential of functional data analysis techniques for analyzing epidemiological data. 
We note that, while some of the techniques employed here are well-established, others are relatively recent and provided new and valuable insights.

%\backmatter

\section*{Acknowledgments}
%\subsection*{Funding}

M.A.~Cremona acknowledges the support of the Natural Sciences and Engineering Research Council of Canada (NSERC), of the Fonds de recherche du Québec Health (FRQS), and of the Faculty of Business Administration, Université Laval. F.~Chiaromonte acknowledges support from the Huck Institutes of the Life Sciences, Penn State.

\subsection*{Author contributions}
All authors conceived ideas and analysis approaches. T.B., J.Di I., and L.T.~retrieved and processed data from multiple public sources, implemented pipelines, and performed statistical analyses. All authors interpreted findings and participated in the writing of the manuscript. M.A.C.~and F.C. supervised the research. 

\subsection*{Data availability}
Data for replication are available at \url{https://github.com/tobiaboschi/fdaCOVID2}.

\bibliography{wileyNJD-AMA}%

\begin{thebibliography}{10}
\providecommand \doibase [0]{http://dx.doi.org/}%

\bibitem{kokoszka2017}
Kokoszka P, Reimherr M. {\it Introduction to functional data analysis}.
\newblock CRC Press .
\newblock 2017.

\bibitem{ramsay2005}
Ramsay JO, Silverman BW. {\it Functional data analysis}.
\newblock Springer.
\newblock 2~ed. 2005.

\bibitem{boschi2021functional}
Boschi T, Di~Iorio J, Testa L, Cremona MA, Chiaromonte F. Functional data
  analysis characterizes the shapes of the first {COVID}-19 epidemic wave in
  {I}taly. {\it Scientific reports} 2021\string; 11(17054)\string: 1--15.

\bibitem{collazos2023modeling}
Collazos JA, Dias R, Medeiros MC. Modeling the evolution of deaths from
  infectious diseases with functional data models: The case of {COVID}-19 in
  {B}razil. {\it Statistics in Medicine} 2023\string; 42\string: 993-1012.

\bibitem{engle2020staying}
Engle S, Stromme J, Zhou A. Staying at home: mobility effects of {COVID}-19.
  {\it Available at SSRN 3565703} 2020\string: 1--16.

\bibitem{nepomuceno2020besides}
Nepomuceno MR, Acosta E, Alburez-Gutierrez D, Aburto JM, Gagnon A, Turra CM.
  Besides population age structure, health and other demographic factors can
  contribute to understanding the {COVID-19} burden. {\it Proceedings of the
  National Academy of Sciences} 2020\string; 117(25)\string: 13881--13883.

\bibitem{nouvellet2021reduction}
Nouvellet P, Bhatia S, Cori A, et al. Reduction in mobility and {COVID}-19
  transmission. {\it Nature communications} 2021\string; 12(1)\string: 1--9.

\bibitem{hastie2009elements}
Hastie T, Tibshirani R, Friedman J. {\it The elements of statistical learning:
  data mining, inference, and prediction}.
\newblock Springer Science \& Business Media .
\newblock 2009.

\bibitem{ciminelli2020covid}
Ciminelli G, Garcia-Mandic{\'o} S. {COVID}-19 in {I}taly: an analysis of death
  registry data. {\it Journal of Public Health} 2020\string; 42(4)\string:
  723--730.

\bibitem{henry2022variation}
Henry NJ, Elagali A, Nguyen M, Chipeta MG, Moore CE. Variation in excess
  all-cause mortality by age, sex, and province during the first wave of the
  {COVID}-19 pandemic in {I}taly. {\it Scientific reports} 2022\string;
  12(1)\string: 1--12.

\bibitem{buoro2020papa}
Buoro S, Di~Marco F, Rizzi M, et al. Papa {G}iovanni {XXIII} {B}ergamo Hospital
  at the time of the {COVID}-19 outbreak: letter from the warfront. {\it J Lab
  Hematol.} 2020\string; 42\string: 8--10.

\bibitem{senni2020covid}
Senni M. {COVID}-19 experience in {B}ergamo, {I}taly. {\it Eur Heart J.} 2020.

\bibitem{golinelli2021small}
Golinelli D, Lenzi J, Adja K, et al. Small-scale spatial analysis shows the
  specular distribution of excess mortality between the first and second wave
  of the {COVID}-19 pandemic in {I}taly. {\it Public Health} 2021\string;
  194\string: 182--184.

\bibitem{perico2021bergamo}
Perico N, Fagiuoli S, Di~Marco F, et al. Bergamo and {COVID}-19: how the dark
  can turn to light. {\it Frontiers in medicine} 2021\string; 8\string: 609440.

\bibitem{vinceti2021association}
Vinceti M, Filippini T, Rothman KJ, Di~Federico S, Orsini N. The association
  between first and second wave {COVID}-19 mortality in {I}taly. {\it BMC
  Public Health} 2021\string; 21(1)\string: 1--9.

\bibitem{basellini2021explaining}
Basellini U, Camarda CG. Explaining regional differences in mortality during
  the first wave of {COVID}-19 in {I}taly. {\it Population Studies}
  2021\string; 76\string: 1--20.

\bibitem{cintia2020relationship}
Cintia P, Pappalardo L, Rinzivillo S, et al. The relationship between human
  mobility and viral transmissibility during the {COVID}-19 epidemics in
  {I}taly. {\it arXiv preprint} 2020\string; arXiv:2006.03141\string: 1--37.

\bibitem{craven1978smoothing}
Craven P, Wahba G. Smoothing noisy data with spline functions. {\it Numerische
  mathematik} 1978\string; 31(4)\string: 377--403.

\bibitem{ramsayfdapackage}
Ramsay JO, Wickham H, Graves S, Hooker G. fda: Functional Data Analysis. {\it R
  package version 2.2-6} 2011.

\bibitem{dpc_province}
DPC . {COVID}-19 dati-province.
  \url{https://github.com/pcm-dpc/COVID-19/tree/master/dati-province};  2020.

\bibitem{dpc_region}
DPC . {COVID}-19 dati-regioni.
  \url{https://github.com/pcm-dpc/COVID-19/tree/master/dati-regioni};  2020.

\bibitem{arpino2020no}
Arpino B, Bordone V, Pasqualini M. No clear association emerges between
  intergenerational relationships and {COVID}-19 fatality rates from
  macro-level analyses. {\it Proceedings of the National Academy of Sciences}
  2020\string; 117(32)\string: 19116--19121.

\bibitem{istat_over65}
ISTAT . Popolazione residente al 1$^{\circ}$ gennaio in tutti i comuni.
  \url{http://dati.istat.it/Index.aspx};  2020.

\bibitem{istat_ASC}
ISTAT . Atlante Statistico dei Comuni. \url{http://asc.istat.it/ASC/};  2020.

\bibitem{hm_hosp}
{Ministry of Health} . Open Data.
  \url{https://www.dati.salute.gov.it/dati/dettaglioDataset.jsp?menu=dati\&idPag=18};
  2021.

\bibitem{min_educ}
{Ministry of Education} . Portale unico dei dati della scuola.
  \url{https://dati.istruzione.it/opendata/opendata/catalogo/\#Scuola};  2021.

\bibitem{stekhoven2012missforest}
Stekhoven DJ, B{\"u}hlmann P. MissForest: non-parametric missing value
  imputation for mixed-type data. {\it Bioinformatics} 2012\string;
  28(1)\string: 112--118.

\bibitem{istat_air}
ISTAT . Tavole dati: Ambiente urbano.
  \url{https://www.istat.it/it/archivio/236912};  2020.

\bibitem{google_mob}
Google . Community mobility reports.
  \url{https://www.google.com/covid19/mobility/};  2021.

\bibitem{modi2021estimating}
Modi C, B{\"o}hm V, Ferraro S, Stein G, Seljak U. Estimating {COVID}-19
  mortality in {I}taly early in the {COVID}-19 pandemic. {\it Nature
  communications} 2021\string; 12(1)\string: 1--9.

\bibitem{borghesi2021lombardy}
Borghesi A, Golemi S, Carapella N, Zigliani A, Farina D, Maroldi R. Lombardy,
  {N}orthern {I}taly: {COVID-19} second wave less severe and deadly than the
  first? {A} preliminary investigation. {\it Infectious Diseases} 2021\string;
  53(5)\string: 370--375.

\bibitem{chirico2021covid}
Chirico F, Nucera G, Szarpak L. {COVID}-19 mortality in {I}taly: The first wave
  was more severe and deadly, but only in {L}ombardy region. {\it Journal of
  Infection} 2021\string; 83(1)\string: e16.

\bibitem{zirilli2022covid}
Zirilli A, Limonti F, Alibrandi A. {COVID}-19 Pandemic Waves in {I}taly: An
  Epidemiological Overview about Infections, Swabs and Death Rates. {\it Open
  Journal of Epidemiology} 2022\string; 12(3)\string: 285--299.

\bibitem{hartigan1975clustering}
Hartigan JA. {\it Clustering algorithms}.
\newblock John Wiley \& Sons, Inc. .
\newblock 1975.

\bibitem{randolph2020herd}
Randolph HE, Barreiro LB. Herd immunity: understanding {COVID}-19. {\it
  Immunity} 2020\string; 52(5)\string: 737--741.

\bibitem{zou2005regularization}
Zou H, Hastie T. Regularization and variable selection via the elastic net.
  {\it Journal of the royal statistical society: series B (statistical
  methodology)} 2005\string; 67(2)\string: 301--320.

\bibitem{friedman2021package}
Friedman J, Hastie T, Tibshirani R, Narasimhan B. glmnet. {\it R package
  version 4.1-3} 2021.

\bibitem{boschi2021highly}
Boschi T, Reimherr M, Chiaromonte F. A Highly-Efficient Group Elastic Net
  Algorithm with an Application to Function-On-Scalar Regression. {\it Advances
  in Neural Information Processing Systems} 2021\string; 34\string: 9264--9277.

\bibitem{cuellar2022assessing}
Cu{\'e}llar L, Torres I, Romero-Severson E, et al. Assessing the impact of
  human mobility to predict regional excess death in {E}cuador. {\it Scientific
  reports} 2022\string; 12(1)\string: 1--12.

\bibitem{rahman2022associations}
Rahman MM, Thill JC. Associations between {COVID}-19 Pandemic, Lockdown
  Measures and Human Mobility: Longitudinal Evidence from 86 Countries. {\it
  International Journal of Environmental Research and Public Health}
  2022\string; 19(12)\string: 7317.

\bibitem{horvath2012inference}
Horv{\'a}th L, Kokoszka P. {\it Inference for functional data with
  applications}. 200.
\newblock Springer Science \& Business Media .
\newblock 2012.

\bibitem{carteni2021role}
Carten{\`\i} A, Di~Francesco L, Henke I, Marino TV, Falanga A. The role of
  public transport during the second {COVID}-19 wave in {I}taly. {\it
  Sustainability} 2021\string; 13(21)\string: 11905.

\bibitem{yu2020veridical}
Yu B, Kumbier K. Veridical data science. {\it Proceedings of the National
  Academy of Sciences} 2020\string; 117(8)\string: 3920--3929.

\bibitem{goldsmith2016refund}
Goldsmith J, Scheipl F, Huang L, et al. refund: Regression with functional
  data. {\it R package version 0.1-16} 2016.

\bibitem{azzolina2022regional}
Azzolina D, Lorenzoni G, Silvestri L, Prosepe I, Berchialla P, Gregori D.
  Regional differences in mortality rates during the {COVID-19} epidemic in
  {I}taly. {\it Disaster Medicine and Public Health Preparedness} 2022\string;
  16(4)\string: 1355--1361.

\bibitem{della2020network}
Della~Rossa F, Salzano D, Di~Meglio A, et al. A network model of {I}taly shows
  that intermittent regional strategies can alleviate the {COVID-19} epidemic.
  {\it Nature communications} 2020\string; 11(1)\string: 5106.

\bibitem{guzzi2020spatio}
Guzzi PH, Tradigo G, Veltri P. Spatio-temporal resource mapping for intensive
  care units at regional level for {COVID-19} emergency in {I}taly. {\it
  International journal of environmental research and public health}
  2020\string; 17(10)\string: 3344.

\end{thebibliography}

\clearpage
\newcommand{\beginsupplement}{
    \setcounter{section}{0}
    \renewcommand{\thesection}{S\arabic{section}}
    \setcounter{equation}{0}
    \renewcommand{\theequation}{S\arabic{equation}}
    \setcounter{table}{0}
    \renewcommand{\thetable}{S\arabic{table}}
    \setcounter{figure}{0}
    \renewcommand{\thefigure}{S\arabic{figure}}
    \newcounter{SIfig}
    \renewcommand{\theSIfig}{S\arabic{SIfig}}}
\beginsupplement
% \appendix
\section*{Supplementary Material}

%% fig: mortality and shifts
\begin{figure}[ht]
%\hspace*{-1.5cm}
\centering
\includegraphics[width=\linewidth]{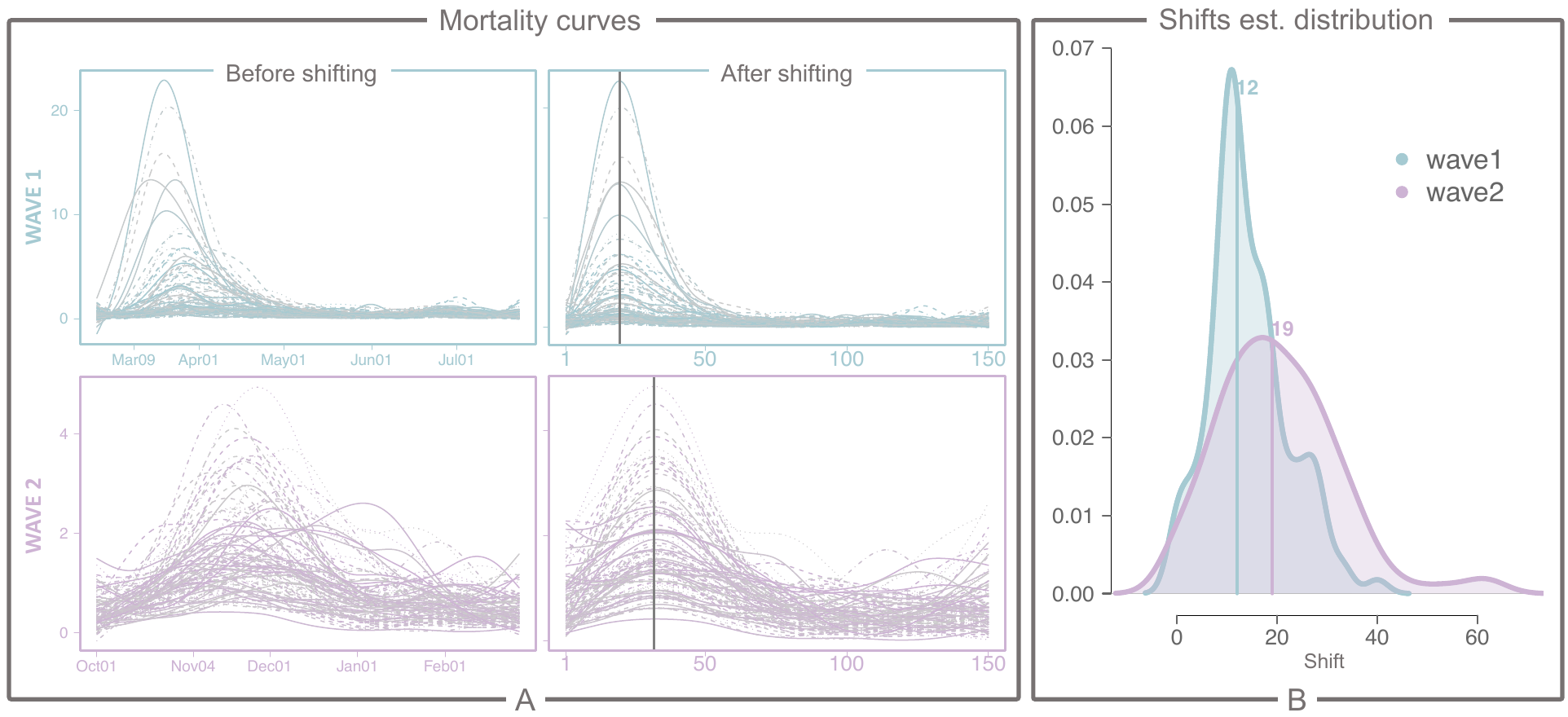}
\caption{
    \textbf{Mortality before and after alignment.} 
    Panel \textbf{A} represents the mortality curves in the first (top) and second (bottom) wave, before (left) and after (right) the shifting to align their peaks. The vertical line on the shifted curves indicates the new common peak time, which has been set to be equal to the time of the earliest peak -- day 20 (peak of Lodi) for the first wave and day 33 (peak of Cagliari) for the second wave. 
    Panel \textbf{B} represents the distributions of the estimated shifts in the two waves. The vertical lines at 12 and 19, respectively, are the median of the two distributions. 
}
\label{supp_fig:mortaliy_and_shifts}
\end{figure}

%% fig: Cluster Dendrograms
\begin{figure}[ht]
\centering
\includegraphics[width=\linewidth]{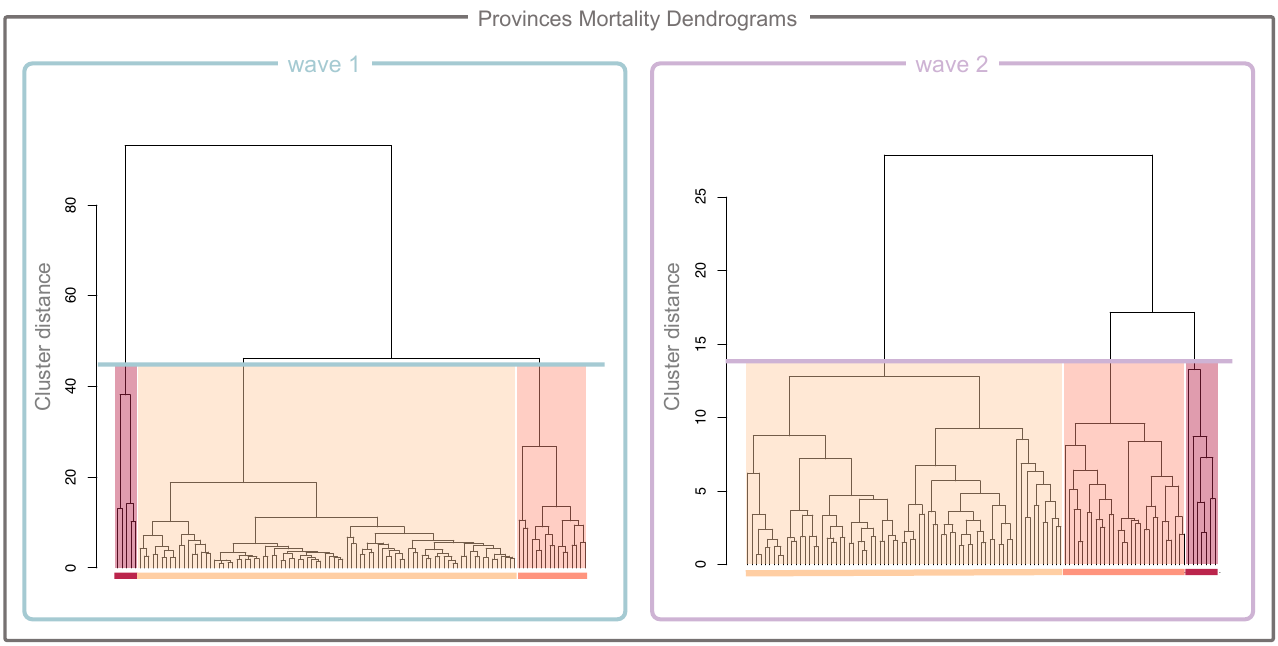}
\caption{
    \textbf{Provinces mortality dendrograms.} The figure displays the dendrograms for both waves, obtained using agglomerative clustering with $L^2$ distance and complete linkage. Note that the vertical axes, representing the distances between clusters, have different scales for the two waves -- implying that clusters in wave 2 are less distinct than the ones in wave 1.
}
\label{supp_fig:provinces_dendrograms}
\end{figure}

%% fig: Cluster membership for provinces
\begin{figure}[ht]
\centering
\includegraphics[width=0.9\linewidth]{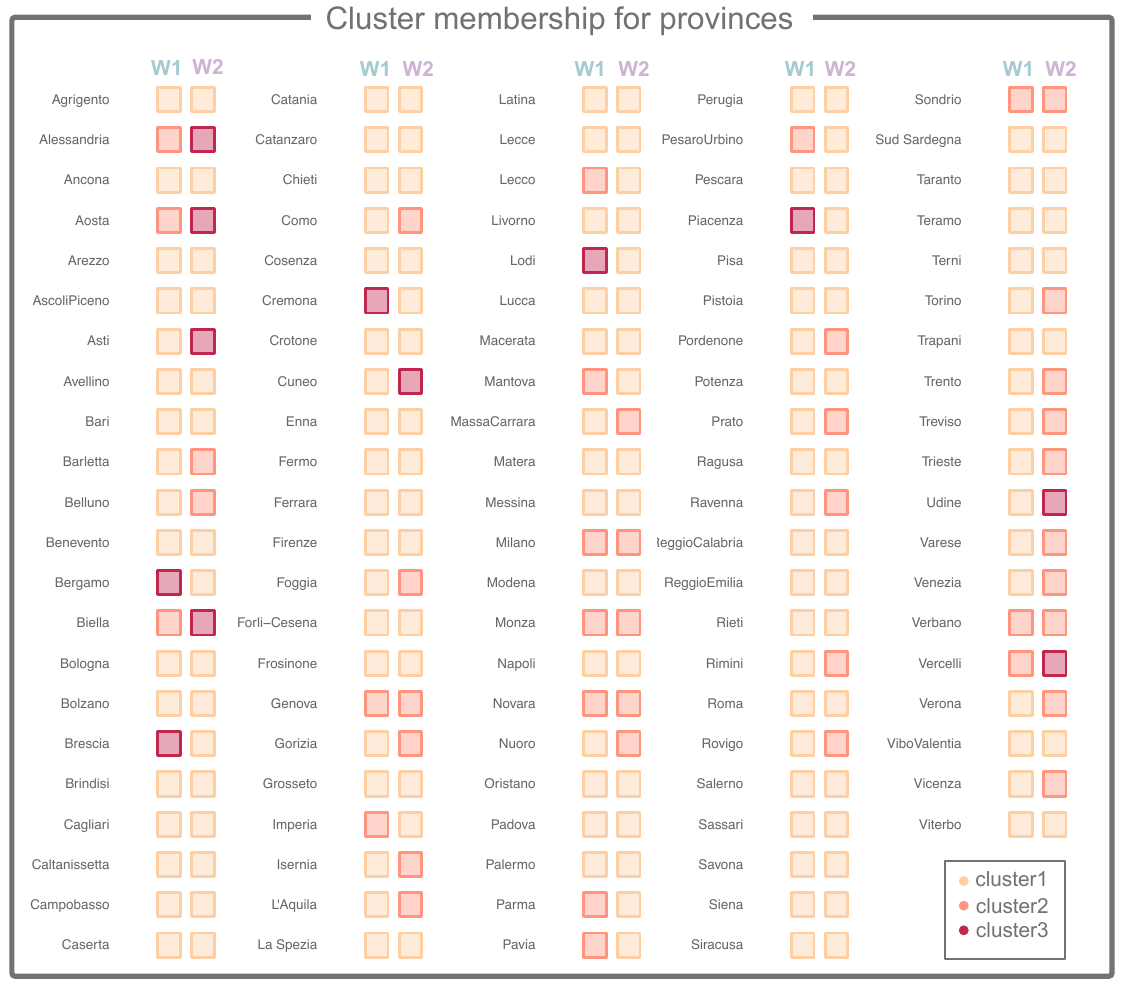}
\caption{
    \textbf{Cluster membership for provinces.} 
    Cluster membership of each province in the first and in the second wave (W1 and W2, respectively).
}
\label{supp_fig:provinces_clusters}
\end{figure}

\clearpage
\begin{table}[t]
    \centering
    \caption{
    \textbf{Joint elastic net fit}. 
    Elastic net coefficients corresponding to optimally chosen $\lambda_1$ via 5-fold cross-validation (we set $\lambda_2 = 0.6 \lambda_1$). For each wave, we show the results corresponding to the minimum cross-validation error ($\lambda_1^{min}$) as well as the ones at one standard deviation of the minimum ($\lambda_1^{1se}$). Models are fitted on the full data set comprising all 107 provinces. All variables are standardized. Dots imply that the associated variable has not been selected by the model (coefficient equal to $0$).}
    \label{tab:elastic_net}
    \begin{tabular}{lllll}
    \toprule
        & \multicolumn{2}{c}{\textbf{Wave 1}} & \multicolumn{2}{c}{\textbf{Wave 2}} \\
        \textbf{Variable} & $\lambda_1^{min} = 0.0036$ & $\lambda_1^{1se} = 0.0978$ & $\lambda_1^{min} = 0.0344$ & $\lambda_1^{1se} = 0.1293$ \\
        \midrule
        \texttt{Over 65} & 0.2079 & 0.1001 & 0.0776 & 0.0106 \\
        \texttt{Adults per family doctor} & 0.0615 & 0.0369 & 0.0035 & $\cdot$  \\
        \texttt{Ave.~beds per hospital} & 0.0778 & 0.0493 & $\cdot$ & $\cdot$ \\
        \texttt{Ave.~students per classroom} & 0.1565 & 0.1037 & 0.0211 & $\cdot$  \\
        \texttt{Ave.~employees per firm} & 0.0595 & $\cdot$ & $\cdot$ & $\cdot$ \\
        \texttt{PM10} & 0.0670 & $\cdot$ & 0.0567 & $\cdot$  \\
        \texttt{Area before} &  0.6049 & 0.5178 & 0.7810 & 0.6649 \\
        \bottomrule
    \end{tabular}
\end{table}

%% fig: joint fgen fit
\begin{figure}[!b]
\centering
\includegraphics[width=0.9\linewidth]{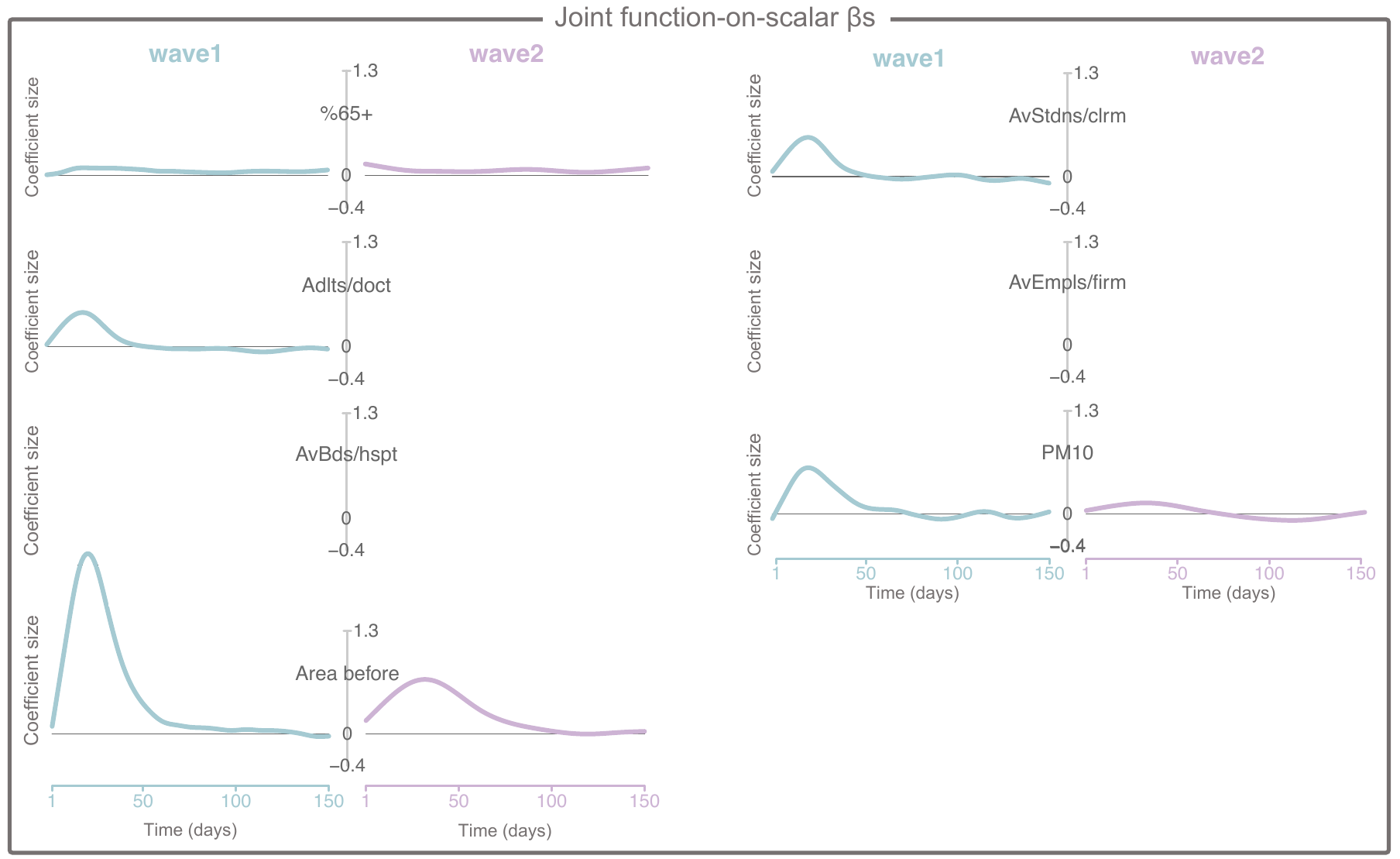}
\caption{
    \textbf{Joint function-on-scalar fit.} 
    \texttt{fgen} functional coefficients corresponding to optimally chosen $\lambda_1$ via 5-fold cross-validation (we set $\lambda_2 = 0.6 \lambda_1$).
    %($\lambda_1$ is equal to $250.9$ for the first wave and to $207.2$ for the second wave). 
    Models are fitted on the full data set comprising all 107 provinces. All variables are standardized. Empty plots imply that the associated variable has not been selected by the model (coefficient equal to $0$).
}
\label{supp_fig:joint_fgen}
\end{figure}

\clearpage
%% fig: scalar beta and pc
\begin{figure}[t]
\centering
\includegraphics[width=0.9\linewidth]{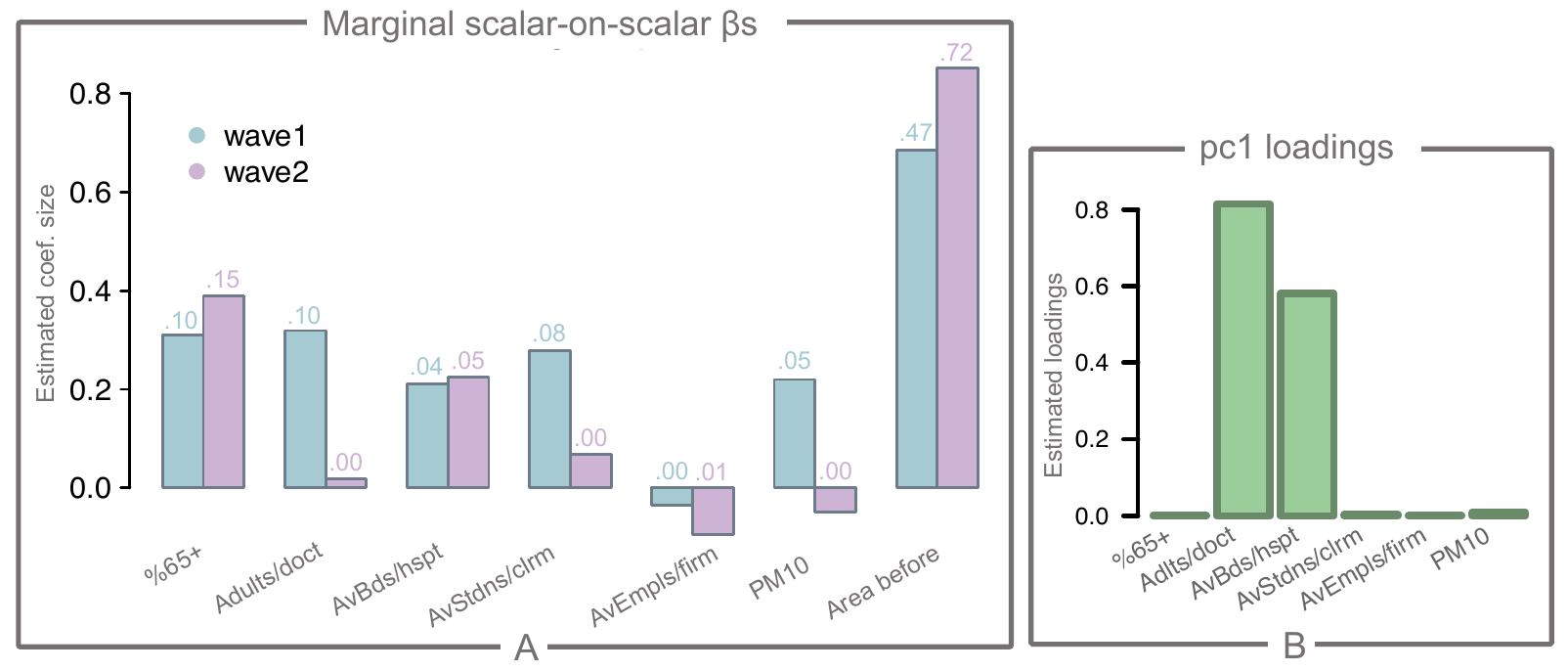}
\caption{
    \textbf{Marginal scalar-on-scalar results.} 
    Panel \textbf{A} shows the estimated coefficients $\hat \beta$ for the marginal regression of $A_{aft}$ (\textit{area after}, scalar response)  on individual scalar features. The associated marginal $R^2$ is reported above each coefficient bar. 
    Panel \textbf{B} shows the loadings of the first principal component computed considering all the scalar features but $A_{bef}$ (which explains $65.6\%$ of the variance).
}
\label{supp_fig:scalar_betas_and_pc1}
\end{figure}

%% fig: function-on-scalar betas
\begin{figure}[b]
%\hspace*{-1.5cm}
\centering
\includegraphics[width=\linewidth]{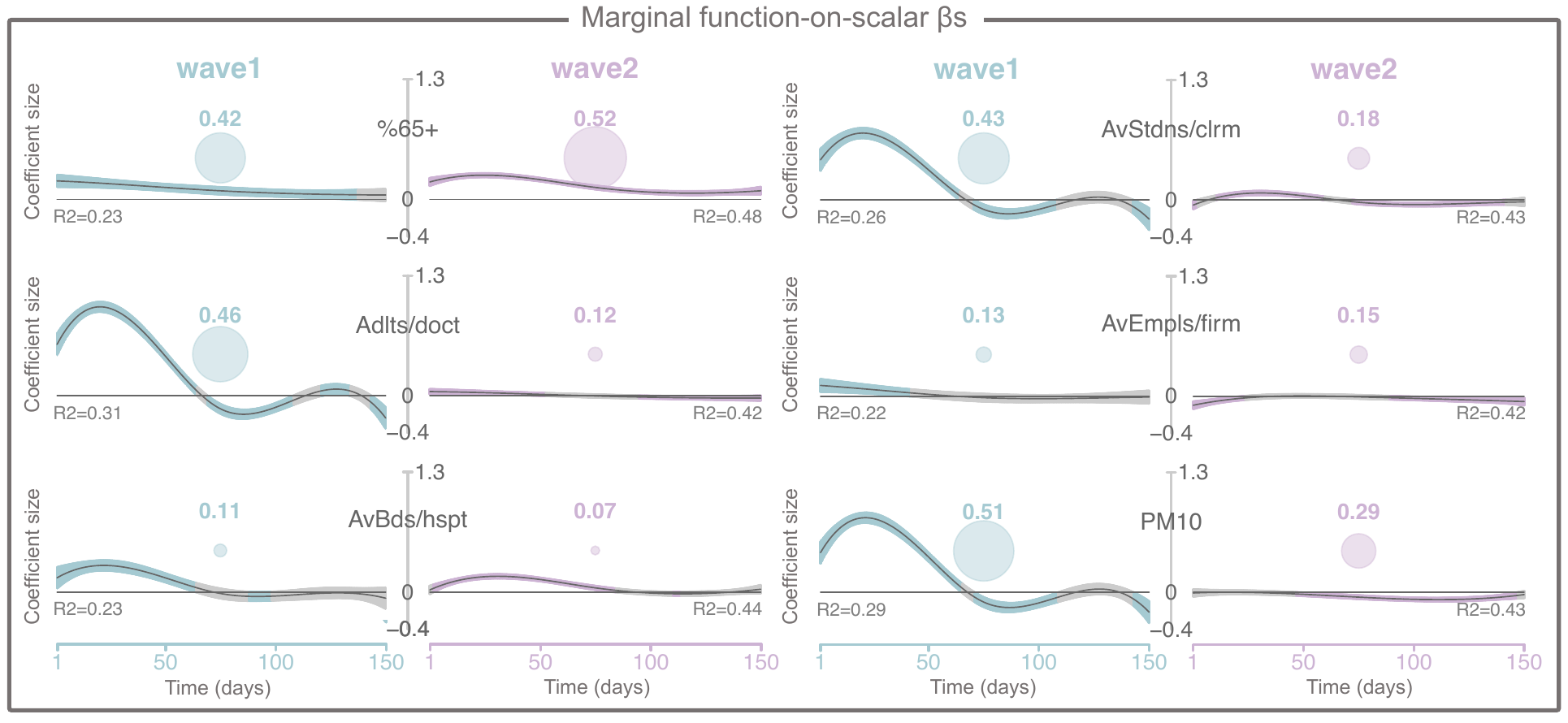}
\caption{
    \textbf{Marginal function-on-scalar results.} 
    The figure depicts the estimated coefficient curves $\hat \beta(t)$ for the marginal regression of the mortality curve on individual scalar covariates. The black solid lines represent $\hat \beta(t)$,  with bands built adding and subtracting $1.96 \times$ pointwise standard errors. Grey areas denote the parts of the time domain where the bands contain the value 0. For each regression, we report its $R^2$ as well as the $\lambda_{max}$-ratio for which the variable enters in the \texttt{fgen} feature selection model (balls are proportional to $\lambda_{max}$-ratio).
    }
\label{supp_fig:function_on_scalar_betas}
\end{figure}

%% fig: heatmap of covariates
\begin{figure}[ht]
\centering
\includegraphics[width=0.7\linewidth]{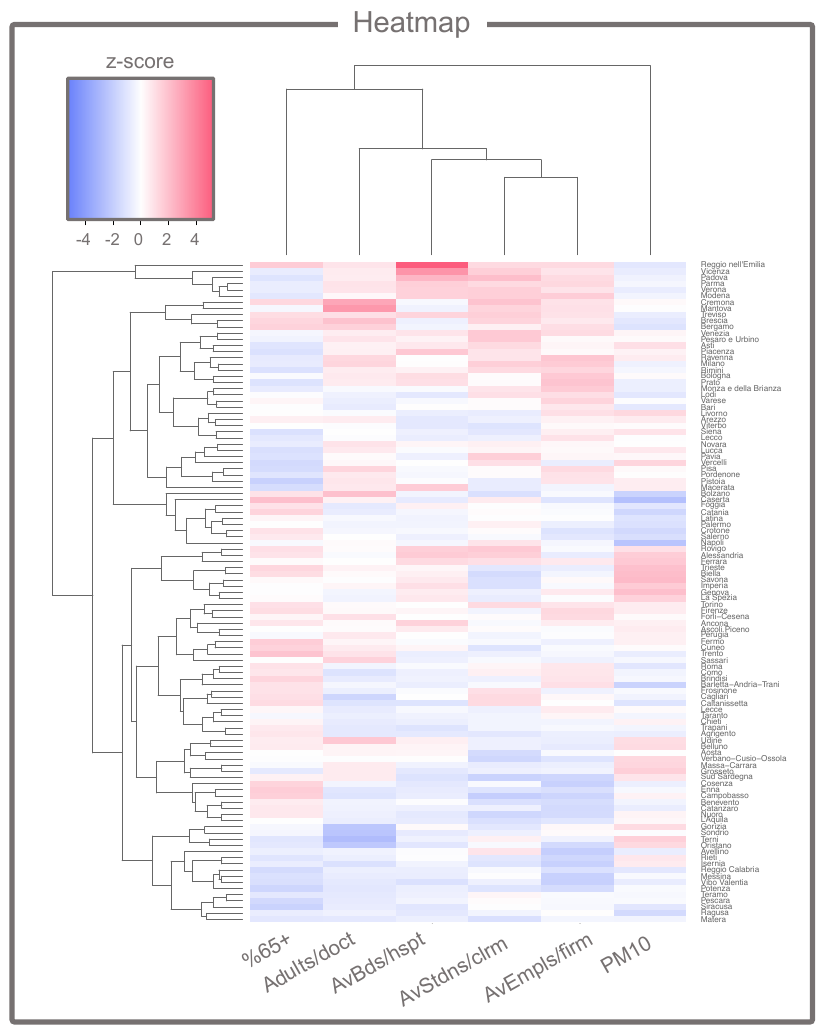}
\caption{
    \textbf{Heatmap of features and provinces.} 
    Heatmap depicting
    %representing, for each province, the magnitude of 
    %its 
    standardized socio-demographic, infrastructural, and environmental features 
    %(i.e., their
    (\textit{z-scores}) across provinces. 
    %Provinces (rows) and
    Features (columns) and provinces (rows) are arranged based on their similarity
    %, obtained 
    -- as captured through hierarchical clustering, and represented by dendrograms (top and left). 
    Absolute correlation distance and complete linkage are employed for agglomerating features;
    %, to capture their covariance structure;
    and Euclidean distance and complete linkage are employed for agglomerating provinces.
    %, to evaluate their grouping structure. 
    %\marzia{[I still don't like the explanation very much, but it's probably just me (and in any case it is not that important]}
}
\label{supp_fig:heatmap}
\end{figure}

%% fig: scalar covariates collinearity
\begin{figure}[ht]
\centering
\includegraphics[width=0.8\linewidth]{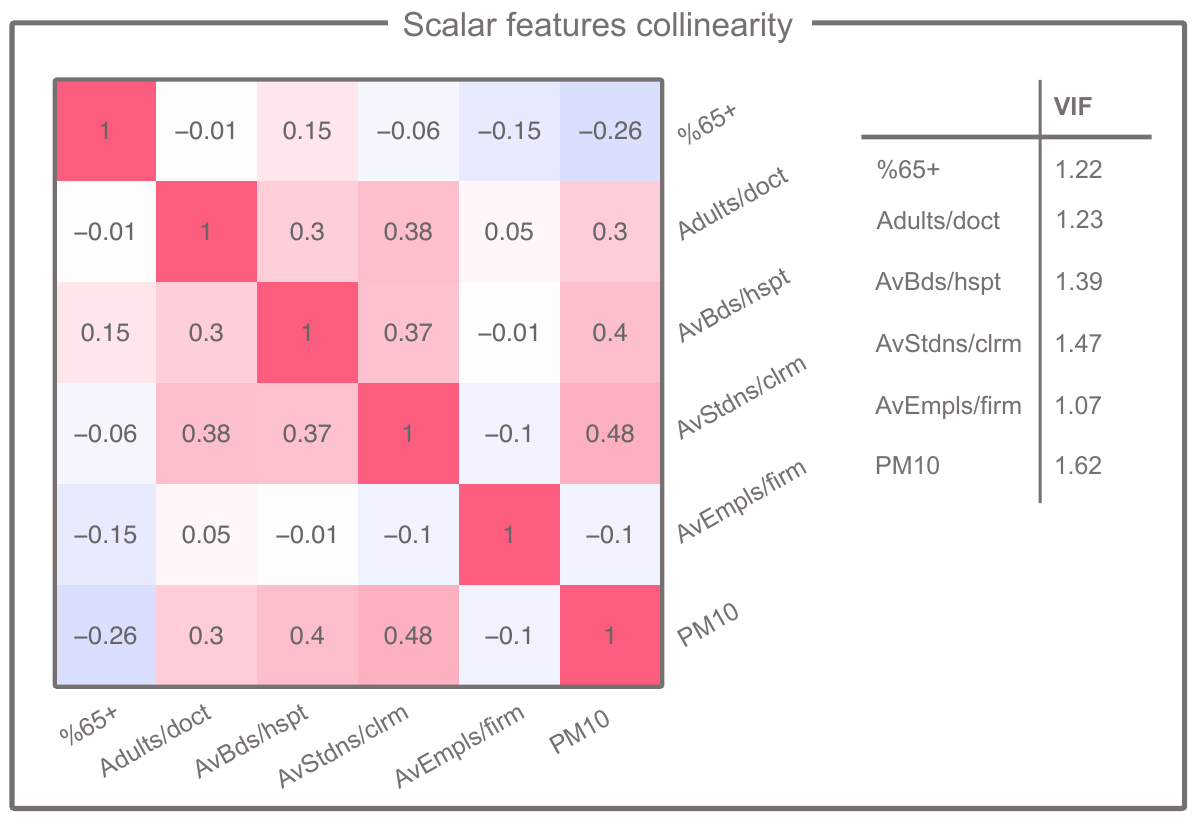}
\caption{
    \textbf{Scalar feature collinearity.} The matrix on the left displays the pairwise correlation between the six scalar features employed in the main analysis. The table on the right contains their variance inflation factors (VIF).
    }
\label{supp_fig:scalar_coll}
\end{figure}

%% fig: sos and fos betas with pc1
\begin{figure}[ht]
%\hspace*{-1.5cm}
\centering
\includegraphics[width=0.9\linewidth]{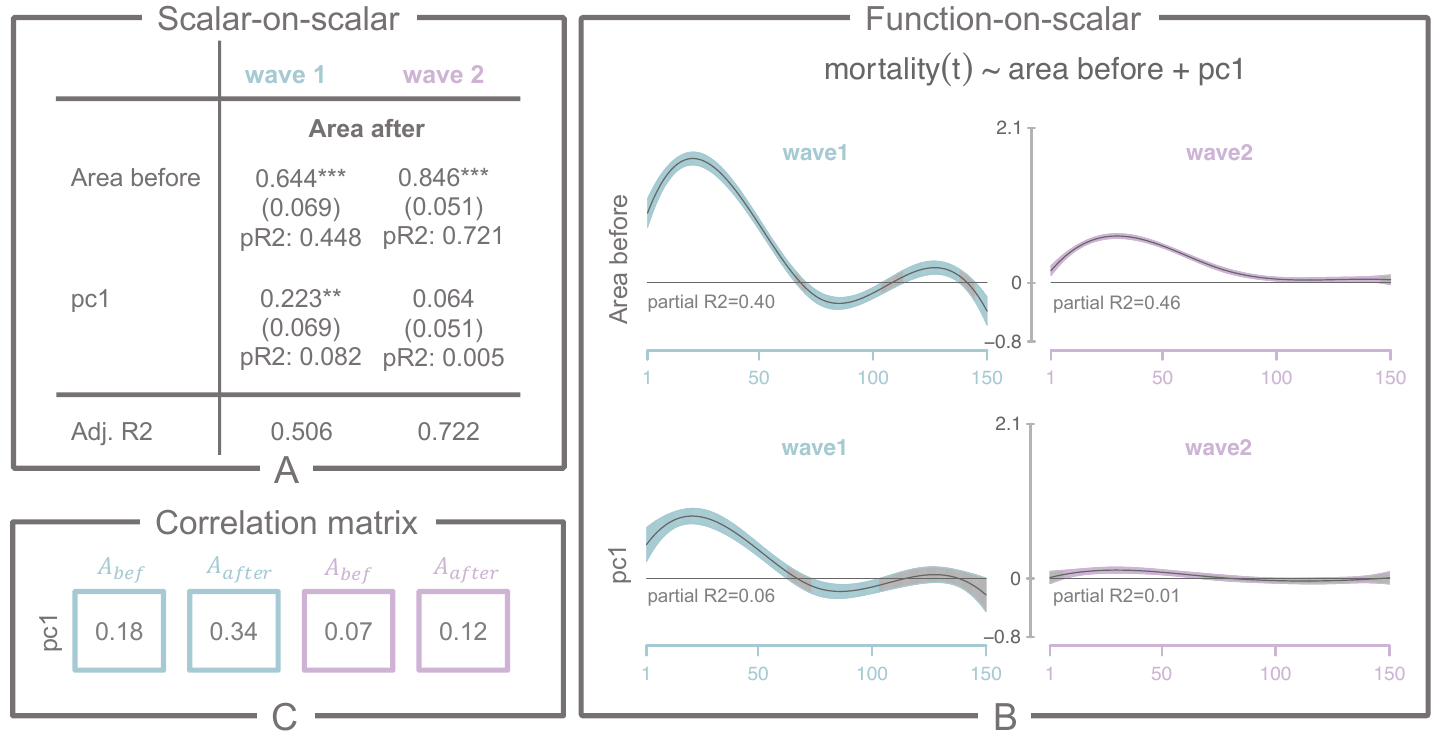}
\caption{
    \textbf{Joint effects of area before and \textit{pc1}.} 
    Panel \textbf{A} shows the results of the joint regression of $A_{aft}$ on $A_{bef}$ and \textit{pc1}. For both waves and for each variable, we show the estimated coefficient (with significance indicated by stars), its standard error in parenthesis, and the associated partial $R^2$. Clearly, area before dominates \textit{pc1} in terms of explanatory power. 
    Panel \textbf{B} depicts the estimated coefficients curves $\hat \beta(t)$ for the joint regression of mortality on the scalar covariates area before and \textit{pc1}. The black solid line represents the estimated mean, with bands built adding and subtracting $1.96 \times$ pointwise standard errors. Grey areas denote the parts of the time domain where the bands contain the value 0. For each regressor, we report the corresponding partial $R^2$. 
    Panel \textbf{C} shows, for each wave, the correlation between $pc1$ and $A_{bef}$ and $A_{after}$.
    }
\label{supp_fig:response_over_areabefore_pc1}
\end{figure}

%% fig: mobility shifted
\begin{figure}[ht]
\centering
\includegraphics[width=0.7\linewidth]{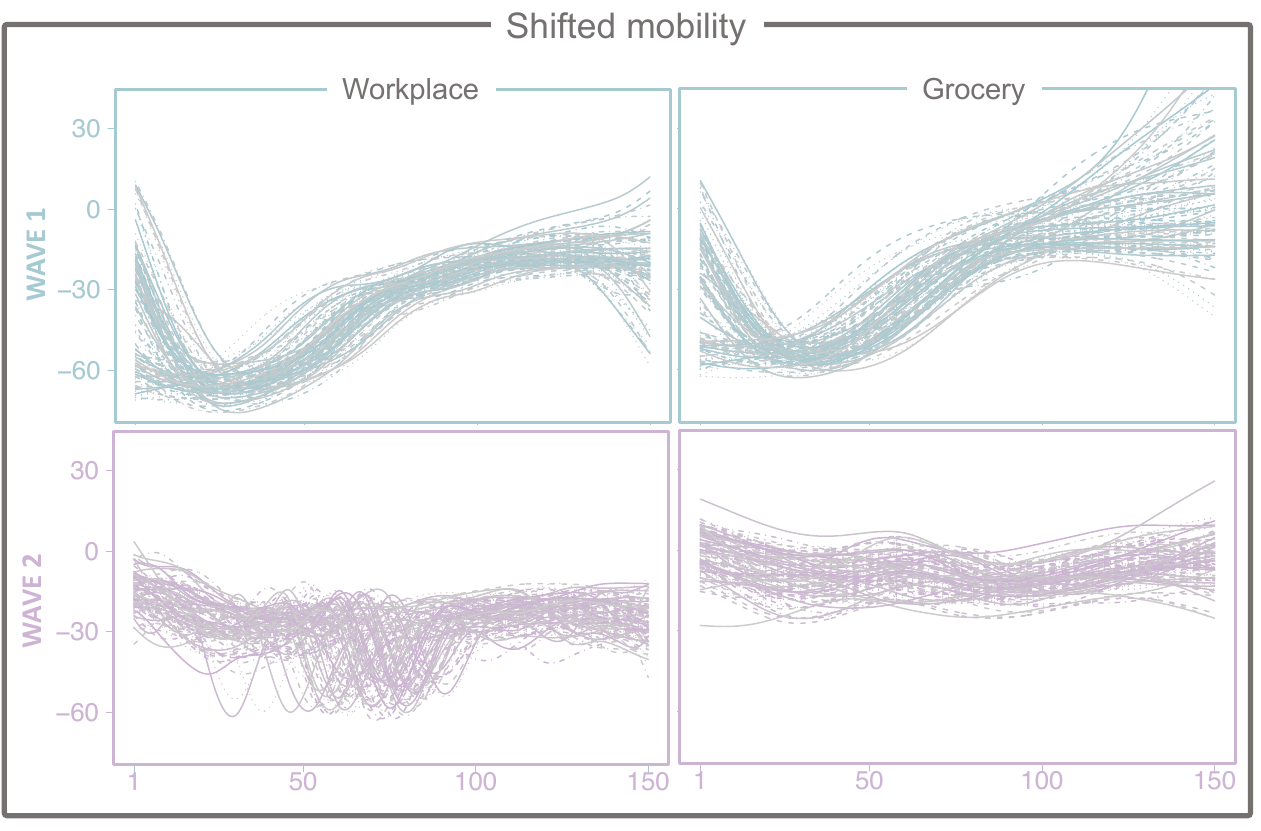}
\caption{
    \textbf{Shifted mobility.} The figure depicts "Workplace" and "Grocery \& Pharmacy" mobility after we apply the shifts estimated on the mortality curves.
    }
\label{supp_fig:mobility_shifted}
\end{figure}

%% fig: function_on_function models
\begin{figure}[ht]
%\hspace*{-1.5cm}
\centering
\includegraphics[width=\linewidth]{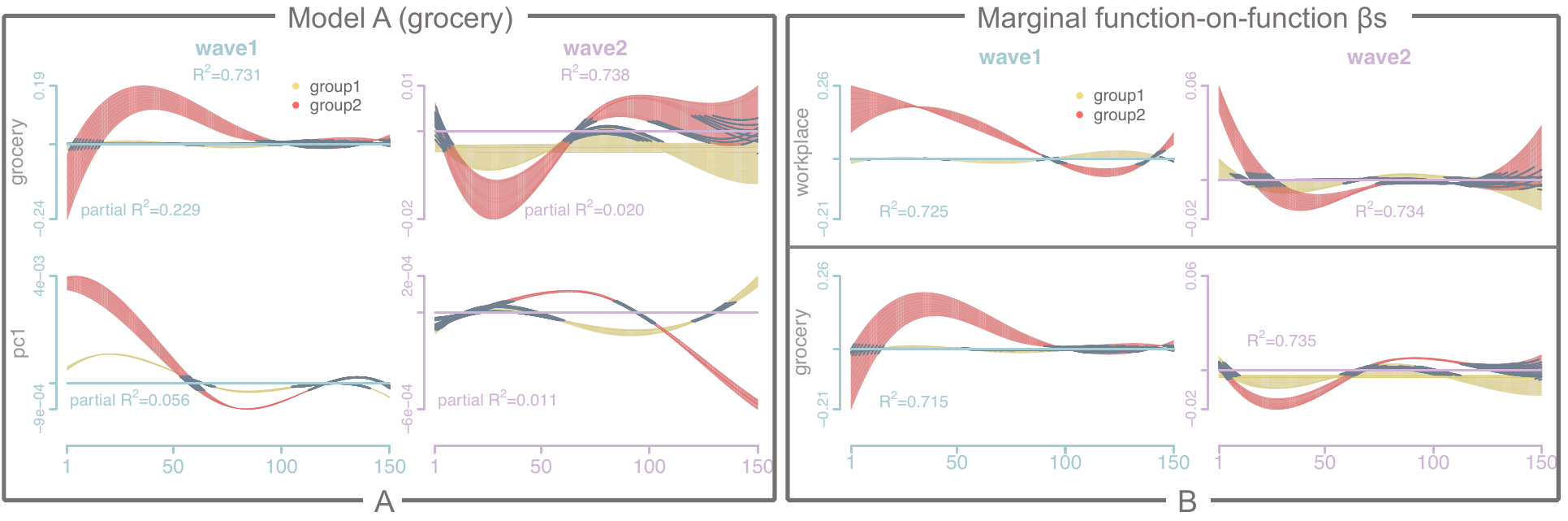}
\caption{
    \textbf{Lagged concurrent functional regression results.} 
    Panel \textbf{A} shows, for both waves, the estimated coefficient curves for the regression of mortality on "Grocery \& Pharmacy" mobility and \emph{pc1}, with an additional binary predictor separating provinces with mild ($d = 0$, Group 1) vs. hard/intermediate ($d = 1$, Group 2) epidemic courses. The plots show ``beams'' comprising 10 curves, one for each of the lags considered in the model ($\ell=15,\dots,24$). Gray portions in a curve correspond to time intervals where it did not significantly depart from $0$ (the confidence interval obtained adding and subtracting to the estimate $1.96 \times$ pointwise standard errors contained $0$). For each regression, we report the average total and partial $R^2$'s over the 10 fits with different lags.
    Panel \textbf{B} shows, in a format similar to {\bf A}, results for the marginal regression of mortality on mobility curves. 
    }
\label{supp_fig:function_on_function_models}
\end{figure}

\end{document}